\documentclass[12pt]{article}

\usepackage{amsmath}
\usepackage{slashed}
\usepackage{amsfonts}
\usepackage{amssymb}
\usepackage{graphicx}
\usepackage{grffile,cite}
\usepackage{subfig}
\usepackage{hyperref}
\usepackage{mathtools}
\usepackage{tikz}
\usepackage{xcolor}

\usepackage{float}

\begin{document}
\thispagestyle{empty}

\begin{center}

{\LARGE\bf{Two loop mass renormalisation in heterotic string theory: NS states}}
\bigskip

{\large Ritabrata Bhattacharya}\, \, \\

\bigskip 
\bigskip

{\small
 Utah State University,
4415 Old Main Hill, Logan, UT - 84322, USA\\[3mm]
}

\end{center}
\begin{center}
\it{ritabrata.bhattacharya@usu.edu}
\end{center}
\bigskip 
 %
%
\begin{center} 
{\bf Abstract} 
\end{center}

\begin{quotation}
\noindent 
{\small

In this work computation of the renormalised mass at two loop order for the NS sector of heterotic string theory is attempted. We first implement the vertical integration prescription for choosing a section avoiding the spurious poles due to the presence of a required number of picture changing operators. As a result the relevant amplitude on genus 2 Riemann surface can be written as a boundary term. We then identify the 1PI region of the moduli space having chosen a gluing compatible local coordinates around the external punctures. We also identify the relevant integrands and the relevant region of integration for the modular parameters at the boundary. }
\end{quotation}

\newpage
\tableofcontents
\setcounter{footnote}{0}
 
\section{Introduction and Summary}
For string theory in critical dimensions conformal invariance plays a crucial role. Demanding conformal invariance we get the on-shell condition for the string states $k_i^2=-m_i^2$ with $m_i$ as the \textit{tree level} mass for the $i$-th mass level of the strings. On the other hand defining S-matrix elements via the LSZ prescription imposes the constraint $k_i^2=-m_{Ri}^2$ where $m_{Ri}$ denotes the renormalized mass of that state. Since conformal invariance implies manifest UV finiteness, hence renormalized masses in string theory are finite unlike in the case of ordinary quantum field theory. For states whose masses are protected against quantum corrections we have $m_{Ri}=m_i$ and S-matrix elements can be defined without any problem. But for generic states in string theory we have $m_{Ri}\neq m_i$ and for them there is an apparent conflict and usual string amplitudes do not compute S-matrix beyond tree level in such cases. 

One needs an off-shell formulation of string theory to resolve this apparent conflict because once we have a definition of the off-shell amplitudes we can then use the standard LSZ prescription to compute S-matrix elements. An ad hoc definition of off-shell amplitudes can be given \cite{Nelson:1988ic,Vafa:1987es,Vafa:1987ea,Cohen:1985sm,Alvarez-Gaume:1987eux,Alvarez-Gaume:1988lei,Polchinski:1988jq}, and has been fully developed in the context of bosonic string theory. Off-shell amplitudes however depend on spurious data encoded in the choice of local coordinates around the punctures due to the external vertex operators. So the result for such amplitudes are ambiguous. In \cite{Pius:2013sca,Pius:2014iaa} it was shown that the renormalized masses and S-matrix elements computed using off-shell amplitudes do not depend on the spurious data provided we restrict the choice of local coordinates to within a special class $-$ those satisfying the requirement of gluing compatibility. Gluing compatibility implies that near the boundary of the moduli space where the punctured Riemann surface $\Sigma$ degenerates to two separate punctured Riemann Surfaces $\Sigma_1$ and $\Sigma_2$ glued at one each of their punctures by standard plumbing fixture prescription, the choice of local coordinate at the external punctures of $\Sigma$ must agree with those induced from the choice of local coordinates at the punctures of $\Sigma_1$ and $\Sigma_2$. Restricting the choice of local coordinates within this class guarantees that the result for all physical quantities are independent of the coordinate choice.

For bosonic string theory one can talk about an underlying string field theory \cite{Witten:1985cc}\cite{Zwiebach:1992ie}. Off-shell amplitudes computed from this field theory in Siegel gauge fall within the general class of off-shell amplitudes discussed in \cite{Nelson:1988ic}, and automatically provides a set of gluing compatible coordinate system \cite{Rastelli:2007gg}. All physical quantities computed from the general system of gluing compatible coordinates thus agree with those computed from string field theory. The generalisation to the case of superstring theory was carried out in \cite{Sen:2015uaa} (see also \cite{Witten:1986qs,Saroja:1992vw,Berkovits:1995ab,Berkovits:2001im,Okawa:2004ii,Berkovits:2004xh,Kroyter:2012ni,Jurco:2013qra,Iimori:2013kha,Erler:2013xta,Kunitomo:2013mq,Kunitomo:2015usa,deLacroix:2017lif,FarooghMoosavian:2019yke}). Computation of on-shell superstring amplitudes using superconformal coordinates on super-Riemann surfaces have been the topic of much research \cite{Witten:2012bh,Belopolsky:1996cy,Belopolsky:1997jz,DHoker:2001kkt,Witten:2012bg,Donagi:2013spo}. The final result in this language is expressed as integrals over supermoduli space of super-Riemann surfaces instead of ordinary moduli space of Riemann surfaces.     

A more practical approach to compute such physical quantities with a proper definition of off-shell amplitudes in superstring theory was presented in \cite{Sen:2014pia} using the formalism involving picture changing operators(PCO) \cite{Friedan:1985ge},\cite{Verlinde:1987sd}. In this formalism the spurious data resides in the choice of local bosonic coordinates and the locations of the PCOs. So the final result is an integral over the ordinary moduli space of Riemann surfaces. Since \cite{Verlinde:1987sd} it is known that the choice of PCO locations correspond to the choice of gauge of the gravitino field. It is impossible to make a global choice of gauge for the gravitino \cite{Donagi:2013spo} and this breakdown of a global gauge choice shows up in the PCO formalism as spurious singularities of the integration measure appearing in a real co-dimension 2 subspace of the moduli space. The practical approach of computing physical quantities is to specify a choice of section avoiding these spurious singularities using the so called ``vertical integration" procedure as described in \cite{Sen:2014pia}. Although actual computation of renormalized mass consistent with this procedure exits only upto one loop i.e. the two point amplitude on the torus \cite{Sen:2013oza,Sen:2016gqt}.

\begin{figure}[h!]
\centering
\includegraphics[scale=0.5]{"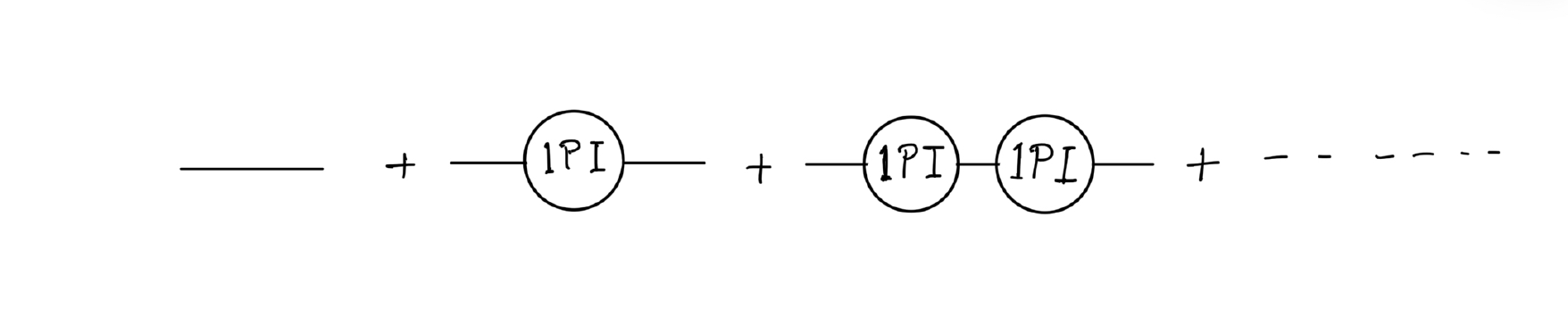"}
\caption{Loop corrected propagator.}
\label{fig:1PI}
\end{figure}

\begin{figure}[h!]
\centering
\includegraphics[scale=0.5]{"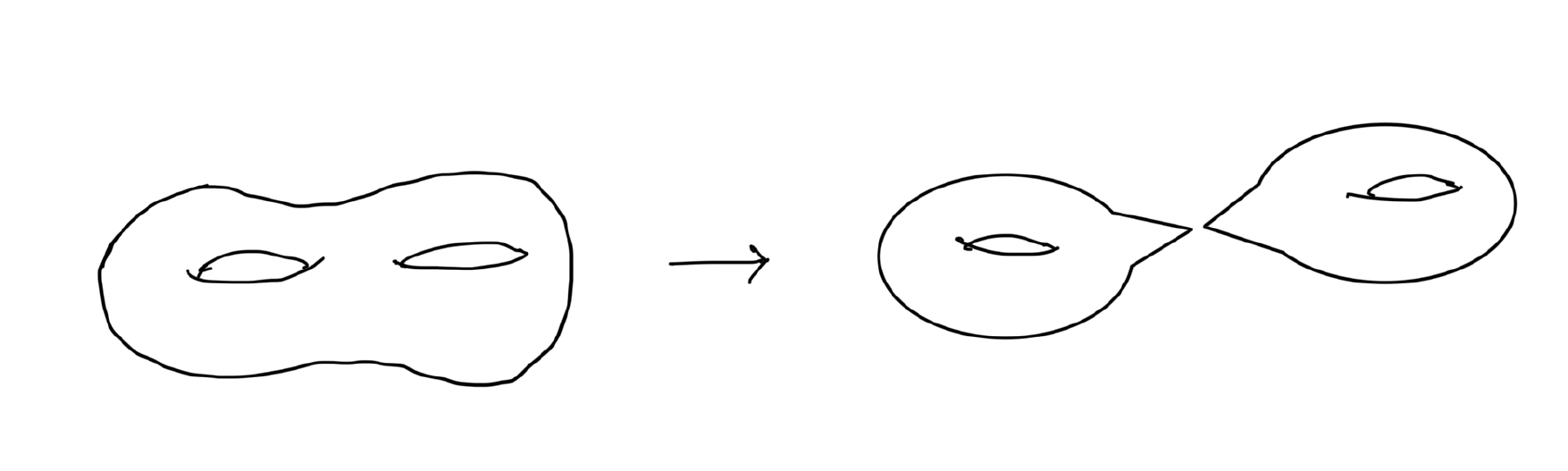"}
\caption{Genus 2 Riemann surface degenerating into two genus 1 surface, each wit 1 extra puncture.}
\label{fig:g2g1}
\end{figure}

\begin{figure}[h!]
\centering
\begin{tikzpicture}
\draw[black, thick] (-5,0) -- (-4.65,0);
\draw[black, thick] (-4.3,0) circle (10pt);
\draw[black, thick] (-3.95,0) -- (-3.6,0);
\draw[black, thick] (-3.25,0) circle (10pt);
\draw[black, thick] (-2.9,0) -- (-2.55,0); 
\draw[black, thick] (-2.15,0) -- (-1.85,0);
\draw[black, thick] (-2,-0.15) -- (-2,0.15);
\draw[black, thick] (-1.45,0) -- (1,0);
\draw[black, thick] (-0.75,0.35) circle (10pt);
\draw[black, thick] (0.3,0.35) circle (10pt);
\end{tikzpicture}
\caption{1 Particle Reducible diagrams at 2 loop order}
\label{fig:1PR}
\end{figure}
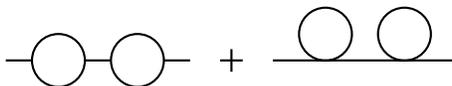

In this paper we attempt to compute the renormalized mass at two loop for heterotic string theory which is relatively simpler to compute. The relevant amplitude one needs to consider in this case is the two point amplitude on a genus 2 Riemann surface. For this one needs to integrate a 10 form over the moduli space of a genus 2 Riemann surface with 2 punctures which has 10 real or 5 complex dimensions. But to get the renormalized mass we must integrate over only a part of the full moduli space which is called the 1 particle irreducible (1PI) subspace in the literature \cite{Pius:2013sca},\cite{Pius:2014iaa}. This follows from our usual intuition in QFT that the loop corrected propagator is obtained by summing over diagrams given by joining 1PI amplitudes by tree level propagators (see figure \ref{fig:1PI}). 
Writing the propagator as $(k^2+m^2)^{-1}$ and the amputated 1PI amplitude as $\mathcal{F}(k)$ we get the loop corrected propagator as,
\begin{equation}
\begin{aligned}
&\frac{1}{k^2+m^2}+\frac{1}{k^2+m^2}\mathcal{F}(k)\frac{1}{k^2+m^2}+\frac{1}{k^2+m^2}\mathcal{F}(k)\frac{1}{k^2+m^2}\mathcal{F}(k)\frac{1}{k^2+m^2}+\dots \\
\Rightarrow\ &\frac{1}{k^2+m^2}\left(1-\frac{\mathcal{F}(k)}{k^2+m^2}\right)^{-1}\quad\Rightarrow\quad\frac{1}{k^2+m^2-\mathcal{F}(k)}
\end{aligned}
\end{equation}
implying that the renormalized mass is given by, $\delta m^2=-\mathcal{F}(k)$. As for the integration over moduli space of genus 2 Riemann surface with two punctures this means we leave out the region where the surface degenerates into two tori each with one extra puncture as in figure \ref{fig:g2g1}.
\\
In terms of Feynman diagrams this amounts to dropping diagrams which are 1 particle reducible as shown in figure \ref{fig:1PR}.
 We will see how to do this explicitly in section \ref{sec:g2p2}. We identify the 1PI subspace explicitly and use the Mapping Class Group (more precisely a subgroup of it) to get the region of integration for the moduli parameters which will be necessary for our case. We will closely follow the method that was used for computing the two loop dilaton tadpole in \cite{Atick:1987qy}. The reason being that in this method starting from the large Hilbert space one writes the relevant amplitude as a total derivative i.e. a boundary term on the moduli space. So the contribution can be identified as coming completely from the boundary of the moduli space. This has several advantages such as the range of integration for the moduli parameters become straight forward as well as identifying the 1PI subspace. The other advantage is more technical in nature since at the boundary we can to deal with the theta functions on the torus rather than those on the genus 2 Riemann surface. Lastly one of the crucial observations of \cite{Atick:1987qy} was that assuming that there is no global obstruction to writing the integrand on the moduli space as a total derivative, the two marked points $p_1$ and $p_2$ of the two tori $T_1$ and $T_2$ in the degeneration limit must approach two of the PCO locations (one on each torus) to keep their locations fixed under global diffeomorphisms of the metric associated with the choice of a gluing compatible local coordinates.

The rest of the paper is organised as follows. In section \ref{sec:review} we set up our conventions and briefly review some results in the literature which will be necessary for our purpose. The results of the ghost and matter conformal field theories are collected in \ref{sec:rev1} and \ref{sec:rev2} respectively. In \ref{sec:vip} we briefly review the so called vertical integration procedure. In section \ref{sec:g2p1} we describe the details of the technique used in \cite{Atick:1987qy} which sets the stage for computing of the two point function. Sections \ref{sec:main}, \ref{sec:main1} and \ref{sec:1pt_sghost} are devoted to the choice of local coordinates, then avoiding the spurious singularities to writing the integrand as a total derivative and then identifying the contribution due to the superghost system. In section \ref{sec:1pt_bcmatter} we add the $b,c$ ghost and matter contribution to show that the 1 point function of the massless states vanish at two loop as expected from the results of \cite{Atick:1986es}. Finally in section \ref{sec:g2p2} we address the mass renormalization of the states in the first massive level of the heterotic string theory. In \ref{sec:g2p2_1} and \ref{sec:g2p2_2} we choose coordinates and avoid the spurious poles as in the 1 point case. Then in \ref{sec:g2p2_3} we identify the 1PI subspace at the boundary of the moduli space and evaluate superghost as well as the $b,c$ ghost contribution which are common for states at all mass levels. Section \ref{sec:massless} shows the vanishing of the renormalised mass for massless states as expected from the non renormalisation theorem \cite{Atick:1986es}. Then section \ref{sec:massive} deals with the renormalised mass for the first massive level and \ref{sec:MCG} gives the region of integration for the moduli parameters. We end with some discussions in section \ref{sec:discuss}.

\section{Conventions and necessary results}
\label{sec:review}
In this section we specify the conventions that we will be using throughout this paper and quote the results that we will need in later sections. Let us emphasise here again that we will be working with the E$_8\times$E$_8$ or SO(32) Heterotic strings to keep our analysis simple. So we have world sheet supersymmetry for only the right moving part hence there are no $\bar{\beta},\bar{\gamma}$ (which are left moving) in this case. All through the paper we set $\alpha'=1$.

\subsection{Ghost CFT}
\label{sec:rev1}
We will use the standard $\xi,\eta,\phi$ CFT to describe the super-conformal $\beta-\gamma$ ghost system as in \cite{Sen:2014pia},
\begin{equation}
\beta=\partial\xi e^{-\phi},\quad \gamma=\eta e^{\phi}\ .
\end{equation}
The (ghost no., picture no., GSO parity) assignment of different ghost fields including the usual conformal ghosts are given by,
\begin{equation}
\begin{aligned}
c,\bar{c}:(1,0,+),\ \ b,\bar{b}:(-1,0,+),\ \ \gamma:(1,0,-),\ \ \beta:(-1,0,-),\\
\xi:(-1,1,+),\ \ \eta:(1,-1,+),\ \ e^{q\phi}:(0,q,(-1)^q)\ .
\end{aligned}
\end{equation}
Thus $e^{\pm\phi}$ are fermionic i.e. they have odd GSO parity. The conformal weights ($\bar{h},h$) of these fields are,
\begin{equation}
\begin{aligned}
\text{Fields}&&\hspace{2cm}\text{Conformal weights}\\
c,\ \bar{c} && (0,-1),\ (-1,0)\\
b,\ \bar{b} && (0,2),\ (2,0)\\
\beta && (0,3/2)\\
\gamma && (0,-1/2)\\
\xi && (0,0)\\
\eta && (0,1)\\
e^{q\phi} && \left(0,-q(q+2)/2\right)
\end{aligned}
\end{equation}\\
Let us now give the relevant OPEs of the different ghost fields (we give the holomorphic ones and the anti-holomorphic ones are exactly same with $\bar{z}$ and $\bar{w}$),
\begin{eqnarray}
&&c(z)b(w)=(z-w)^{-1}+\dots,\quad \xi(z)\eta(w)=(z-w)^{-1}+\dots,\nonumber\\
&& \partial\phi(z)\partial\phi(w)=-(z-w)^{-2}+\dots,\nonumber\\  
&&e^{q_1\phi}(z)e^{q_2\phi}(w)=(z-w)^{-q_1q_2}e^{(q_1+q_2)\phi}(w)+\dots,\\
&\Rightarrow&\quad \beta(z)\gamma(w)=-(z-w)^{-1}+\dots\ .
\end{eqnarray}
The $\dots$ denote terms which are regular. The BRST charge, $Q_B$ is given by,
\begin{equation}
Q_B=\oint dz\ j_B(z)\ +\ \oint d\bar{z}\ \bar{j}_B(\bar{z})\ ,
\end{equation}
with,
\begin{eqnarray}
\bar{j}_B(\bar{z})&=&\bar{c}(\bar{z})\bar{T}_m(\bar{z})+\bar{b}(\bar{z})\bar{c}(\bar{z})\bar{\partial}\bar{c}(\bar{z})\\
j_B(z)&=&c(z)(T_m(z)+T_{\beta,\gamma}(z))+\gamma(z)T_F(z)-b(z)c(z)\partial c(z)-\frac{1}{4}\gamma(z)^2b(z)
\end{eqnarray}
where $T_m$ and $T_{\beta,\gamma}$ are the world-sheet matter and the $\beta,\gamma$ energy momentum tensors respectively. $T_F$ is the world-sheet matter super-current. With these definitions at our disposal we now define the Picture Changing Operator (PCO) which in terms of the $\xi,\eta,\phi$ fields is given by,
\begin{equation}
\chi(z)=\lbrace Q_B,\xi(z)\rbrace=c\partial\xi(z)+e^{\phi}T_F(z)-\frac{1}{4}\partial\eta e^{2\phi}b(z)-\frac{1}{4}\partial(\eta e^{2\phi}b(z))\ .
\label{eq:PCO}
\end{equation}
It is a dimension zero primary operator with picture number 1 satisfying $[Q_B,\chi(z)]=0$. 

The correlation function of $\xi,\eta,\phi$ in the large Hilbert space on a genus $g$ Riemann surface can be found in \cite{Verlinde:1987sd} and is given by,
\begin{equation}
\begin{aligned}
&\bigg\langle\prod_{i=1}^{n+1}\xi(x_i)\prod_{j=1}^{n}\eta(y_j)\prod_{i=1}^{m}e^{q_k\phi(z_k)}\bigg\rangle_{\delta}\\
&=\frac{\prod_{j=1}^{n}\vartheta[\delta]\left(-\vec{y}_j+\sum\vec{x}-\sum\vec{y}+\sum q\vec{z}-2\vec{\Delta}\right)}{\prod_{i=1}^{n+1}\vartheta[\delta]\left(-x_i+\sum\vec{x}-\sum\vec{y}+\sum q\vec{z}-2\vec{\Delta}\right)}\frac{\prod_{i<i'}E(x_i,x_{i'})\prod_{j<j'}E(y_j,y_{j'})}{\prod_{i,j}E(x_i,y_j)\prod_{k<l}E(z_k,z_l)^{q_kq_l}\prod_k\sigma(z_k)^{2q_k}}\ ,\\
&\text{with}\ \sum_{k=1}^m q_k=2(g-1)\ .
\end{aligned}
\label{eq:largeHilbert}
\end{equation}
Here $\vartheta[\delta]$ denotes the theta functions on higher genus Riemann surface with spin structure $\delta$\footnote{\textit{e.g.} For genus $g=1$ they are the elliptic theta functions on the torus.}. The function $E(x,y)$ is called the prime form which has a zero only at $x=y$ and behaves like $E(x,y)\sim (x-y)$ in the limit $x\rightarrow y$. $\sigma(z)$ is a $\frac{1}{2}g$ differential with no zeros or poles, representing the conformal anomaly of the ghost system and $\vec{\Delta}$ is the Riemann class characterising the divisor of zeros of the theta function. For a more elaborate definition of these quantities we refer the reader to \cite{Verlinde:1987sd},\cite{Verlinde:1986kw}. $\sum \vec{x},\ \sum\vec{y}$ and $\sum q\vec{z}$ denote respectively $\sum_{i=1}^{n+1} \vec{x}_i,\ \sum_{j=1}^{n}\vec{y}_j$ and $\sum_{k=1}^{m} q_k\vec{z}_k$ with
\begin{equation}
\vec{x}\equiv\int_p^x\vec{\omega}\ ,
\label{eq:divisor}
\end{equation}
where $\vec{\omega}$ is a $g$-dimensional vector of holomorphic one forms on the Riemann surface(also called abelian differentials) and $p$ is an arbitrary point on the Riemann surface and this dependence on $p$ is compensated by the $p$-dependence of $\vec{\Delta}$.

The fact that $\eqref{eq:largeHilbert}$ is a correlation function on the large Hilbert space is corroborated by the fact that we have one extra number of $\xi$ than the number of $\eta$\footnote{In the large Hilbert space there is a $\xi$ zero mode on any Riemann surface that needs to be soaked up.}. Otherwise this correlation function vanishes.

\textbf{Note}: To get the correlation functions in the small Hilbert space we need to consider a correlation function in the large Hilbert space where all but one $\xi$ has a derivative acting on them. Then we have the following result,
$$\langle\xi(z_0)(\dots)\rangle_{\text{large}}\rightarrow \text{independent of $z_0$.}$$
Thus we have the following prescription for obtaining the correlation function on the small Hilbert space,
\begin{equation}
\langle\xi(z_0)(\dots)\rangle_{\text{large}}\equiv \langle(\dots)\rangle_{\text{small}}
\end{equation} 

\subsection{Matter CFT }
\label{sec:rev2}
\label{sec:CFTmatter}
We know that superstring theory in 10 dimensions is described by the ghost and matter CFT together with total central charge being 0. So along with the ghost CFT described in the previous subsection we also need the matter sector which we describe now. 

The primary operators of the matter CFT in the heterotic string are, $\partial X^{\mu},\bar{\partial}X^{\mu},\psi^{\mu}\ (\mu=0,...,9)$ and $e^{ik.X}$\footnote{We also have $\bar{\psi}^{\mu}$ along with the others for the type II string theories.} with conformal weights $(\bar{h},h)$ given by,
\begin{equation}
\begin{aligned}
\text{Fields}&&\hspace{2cm}\text{Conformal weights}\\
\partial X^{\mu},\ \bar{\partial}X^{\mu} && (0,1),\ (1,0)\\
\psi^{\mu} && (0,1/2)\\
e^{ik.X} && \left(k^2/4,k^2/4\right)\\
S_1^{\pm},\dots, S_5^{\pm} && (1/8,0)
\end{aligned}
\end{equation}
All the matter fields have ghost no.= picture no.= 0. As for the GSO parity, only $\psi^{\mu}$ are GSO odd($-$), all other fields are GSO even($+$). Finally we need the OPE of these fields and we provide the ones which are singular (again we give only the holomorphic ones),
\begin{eqnarray}
\partial X^{\mu}(z)\partial X^{\nu}(w)\ =\ -\frac{\eta^{\mu\nu}}{2(z-w)^2}+\dots ,&&\psi^{\mu}(z)\psi^{\nu}(w)\ =\ \frac{\eta^{\mu\nu}}{2(z-w)}+\dots\ ,\\
\partial X^{\mu}(z)e^{ik.X(w,\bar{w})}&=&-\frac{ik^{\mu}}{2(z-w)}e^{ik.X(w,\bar{w})}+\dots\ .
\end{eqnarray}
The factor of $1/2$ in the $\psi,\psi$ OPE is taken so that the complex fermions built out of the Majorana ones, have the OPE $\Psi_i(z)\bar{\Psi}_i(w)\sim(z-w)^{-1}$. In \cite{Polchinski:1998rr} no such factor is there so $\Psi_i,\bar{\Psi}_i$ OPE will have a factor of 2. Using bosonization we can write\footnote{Here we do the usual Wick rotation $\psi^0\rightarrow i\psi^{10}$ to define the complex fermions appropriately.},
$$\Psi_j=\frac{1}{\sqrt{2}}(\psi^{j}+i\psi^{j+5})\equiv e^{iH_j},\quad\bar{\Psi}_j=\frac{1}{\sqrt{2}}(\psi^{j}-i\psi^{j+5})\equiv e^{-iH_j},\ (j=1,\dots,5)$$
$$S_j^{\pm}\equiv e^{\pm \frac{iH_j}{2}},\quad :\bar{\Psi}_i\Psi_i:\ \equiv \partial H_i\ .$$
We also have,
\begin{equation}
\begin{aligned}
&\bar{\Psi}_i(z)S_i^+(w)\sim (z-w)^{-\frac{1}{2}}S_i^-\ ,\quad \Psi_i(z)S_i^-(w)\sim (z-w)^{-\frac{1}{2}}S_i^+\ ,\\
&\bar{\Psi}_i(z)S_i^-(w)\sim (z-w)^{\frac{1}{2}}\hat{S}_i^-\ ,\quad \Psi_i(z)S_i^+(w)\sim (z-w)^{\frac{1}{2}}\hat{S}_i^+\ ,\\
&\bar{\Psi}_i(z)\Psi_i(w)\sim (z-w)^{-1}\ ,\quad S_i^+(z)S_i^-(w)\sim (z-w)^{-\frac{1}{4}}\ .
\end{aligned}
\end{equation}
All the other ones are regular.
Let us also give the expression of $T_F$ here as it will be needed for later analysis.
\begin{equation}
T_F(z)=i\sqrt{2}\ \psi_{\rho}\partial X^{\rho}\ .
\label{eq:supercurrent}
\end{equation}

For the case of genus 2, the correlation function of the different chiral matter fields were provided in \cite{DHoker:2001kkt} and the result for non-chiral $X^{\mu}$ can be found in \cite{Verlinde:1987sd}. Let us first state some necessary definitions.
\begin{itemize}
\item{Prime form :-
\begin{equation}
E(x,y)=\frac{\vartheta[\delta_{1}](\int_y^x\vec{\omega}|\Omega)}{\zeta(x)^{1/2}\zeta(y)^{1/2}}\sim(x-y)dx^{-1/2}dy^{-1/2}\ (\text{for $x\sim y$}),
\label{eq:prime_form}
\end{equation}
where, $\delta_{1}$ denotes the odd spin structure i.e. $\vartheta[\delta_{1}](0|\Omega)=0,\ \partial_{z_i}\vartheta[\delta_{1}](0|\Omega)\neq 0$. $\zeta(x)$ is given by,
$$\sum_{i=1}^2\partial_{z_i}\vartheta[\delta_{1}](0|\Omega)\omega_i(x).$$ 
If we label the the four non contractible cycles of the genus 2 surface by, $a_1,b_1$ and $a_2,b_2$ then,
$$\int_{a_i}\omega_j=\delta_{ij},\quad \Omega_{ij}=\int_{b_i}\omega_j\ ,\ \ \forall i=1,2$$ defines the period matrix $\Omega$ with a fixed normalisation.}
\item{Szego Kernel :-
\begin{equation}
S_\delta(x,y)=\frac{\vartheta[\delta](\int_y^x\vec{\omega}|\Omega)}{\vartheta[\delta](0|\Omega)E(x,y)}\sim\frac{1}{x-y}dx^{1/2}dy^{1/2}\ (\text{for $x\sim y$}).
\end{equation}
Here $\delta$ denotes the even spin structures i.e. $\vartheta[\delta](0|\Omega)\neq 0$.}
\end{itemize}
With these definitions we have the following correlation functions on genus 2 for the matter sectors,
\begin{eqnarray}
\langle X^{\mu}(z_1,\bar{z}_1)X^{\nu}(z_2,\bar{z}_2)\rangle_{p_k}&=&-\frac{\eta^{\mu\nu}}{2}\ln|E(z_1,z_2)|^2+2\pi p^{\mu}.\text{Im}(\vec{z}_{12}) p^{\nu}.\text{Im}(\vec{z}_{12})\ ,\label{eq:S_cor}\\
\langle\psi^{\mu}(z_1)\psi^{\nu}(z_2)\rangle&=&-\frac{1}{2}S_{\delta}(z_1,z_2)\ ,\label{eq:F_cor}\\
\left\langle\prod_j :e^{ik_j.X(z_j,\bar{z}_j)}:\right\rangle&=&C^X_{g=2}(2\pi)^{10}\delta^{10}\left(\sum_jk_j\right)\prod_{i<j}|E(z_i,z_j)|^{k_i.k_j}e^{-2\pi k_i^{\mu}p_{\mu}.\text{Im}(\vec{z}_{ij}) k_j^{\nu}p_{\nu}.\text{Im}(\vec{z}_{ij})}\nonumber\\
\label{eq:V_cor}
\end{eqnarray}
where $p_k^{\mu}$ denotes the loop momenta along the $b_k$ cycle with $k=1,2$ for a genus 2 surface. They are related to the determinant of imaginary part of the period matrix. Also we have,
$$\text{Im}(\vec{z}_{ij})=\frac{1}{2i}\left(\int_{z_i}^{z_j}\vec{\omega}-\int_{\bar{z}_i}^{\bar{z}_j}\vec{\bar{\omega}}\right)\ \ \Rightarrow\ \ p^{\mu}.\text{Im}(\vec{z}_{ij})\equiv\frac{1}{2i}\sum_{k=1}^2\left(\int_{z_i}^{z_j}p_k^{\mu}\omega_k-\int_{\bar{z}_i}^{\bar{z}_j}p_k^{\mu}\bar{\omega}_k\right)\ .$$
Finally $C^X_{g=2}$ is the normalisation due to the matter partition function and the $\partial_z E(0)$ in the denominator is due to the self contractions. The correlation function for the vertex fields can be found by generalising the torus result given in \cite{Polchinski:1998rr}.
With these, all other correlation functions (for the NS states)\footnote{For R sector we also need the correlation function for the spin fields which we do not discuss.} can be determined as in the case of torus, \cite{Sen:2013oza,Sen:2016gqt,Atick:1986rs}.

\subsection{Vertex operators for the Heterotic string}
Let us now consider the full matter ghost CFT and define a subspace $\mathcal{H}_0$ of the off-shell NS string states with picture number -1 by,
\begin{equation}
|\Psi\rangle\in\mathcal{H}_0\ \ \text{if}\ \ (b_0-\bar{b}_0)|\Psi\rangle=0,\ \ (L_0-\bar{L}_0)|\Psi\rangle=0,\ \ \eta_0|\Psi\rangle=0,\ \ \text{Picture no.}(|\Psi\rangle)=-1 .
\end{equation}
The $\eta_0|\Psi\rangle=0$ condition implies that we are working in the small Hilbert space \cite{Friedan:1985ge}. Let us also define $\mathcal{H}_1\subset\mathcal{H}_0$ containing off-shell states with ghost number 2 in the Siegel gauge,
\begin{equation}
|\Psi\rangle\in\mathcal{H}_1\ \ \text{if}\ \ |\Psi\rangle\in\mathcal{H}_0,\ \ (b_0+\bar{b}_0)|\Psi\rangle=0,\ \ \ \text{Ghost no.}(|\Psi\rangle)=2 .
\end{equation}
Since our ultimate goal is to compute the renormalized mass of the on-shell string states, let us give the on-shell vertex operators for the E$_8\times$E$_8$ or SO(32) Heterotic theory, which for non-zero momentum has the general form,
\begin{equation}
V_i(z,\bar{z})=c\bar{c}e^{-\phi}W_i(z,\bar{z}),\ \ \Rightarrow\ \ \text{On-shell states}:=\ \ |V_i\rangle=V_i(0,0)|0\rangle\in\mathcal{H}_1
\end{equation}
where $W_i(z,\bar{z})$ are some dimension (1,1/2) super-conformal primary with odd GSO parity so that $V_i(z,\bar{z})$ is GSO even. By $|0\rangle$ we denote the SL(2,$\mathbb{C}$) invariant vacuum of the CFT. 

For concreteness let us specify the CFT with central charge 16 for the anti-holomorphic part which does have world-sheet supersymmetry.
\begin{itemize}
\item{We introduce 16 scalars $\mathcal{Y}^I(\bar{z})$ with $I=1,\dots,16$}
\item{We consider theory with weight (1,0) primary fields given by,
\begin{eqnarray}
\bar{J}^a(\bar{z})&=&\bar{\partial}\mathcal{Y}^I\ \text{for $a,I=1,\dots,16$}\\
\bar{J}^a(\bar{z})&=& e^{i\vec{m}.\vec{\mathcal{Y}}}\ \text{with $\vec{m}.\vec{m}=4$, for $a=17,\dots,496$.}
\end{eqnarray}
Note in the above, that $\vec{m}$ is an array of 16 numbers taking value in the dual lattice of E$_8\times$E$_8$ or SO(32). As a result $e^{i\vec{m}.\vec{\mathcal{Y}}}$ with $\vec{m}.\vec{m}=4$ gives us a total of 480 operators. So all total we have 496 operators which is exactly equal to the number of generators of E$_8\times$E$_8$ or SO(32) symmetry group.}
\end{itemize}
This specifies the field content of the heterotic strings and in turn fixes the partition function.

\subsection{Vertical integration procedure}
\label{sec:vip}
Unlike the case in bosonic string theory, an additional subtlety that one has to take into account while computing amplitudes in superstring theory is the presence of spurious poles (i.e. divergences occurring even when no two external vertices are coming close to each other) which can occur in the following cases:
\begin{itemize}
\item{The PCOs colliding with the vertex operators.}
\item{Two or more PCOs colliding with each other.}
\item{In case of higher genus surfaces there some special points where the theta functions vanish even if no operators collide with each other. For example, in equation \eqref{eq:largeHilbert} the $\vartheta[\delta]$ has a divisor of $g$ zeros due to the Riemann vanishing theorem so whenever the denominator vanishes we get a spurious pole.}
\end{itemize}
In the language of super-moduli space the source of these poles are the breakdown of supersymmetry gauge choices in different regions of the moduli space. The remedy then, is to make different gauge choice in different patches and appropriate matching at the boundary. 

\begin{figure}[h!]
\centering
\includegraphics[scale=0.5]{"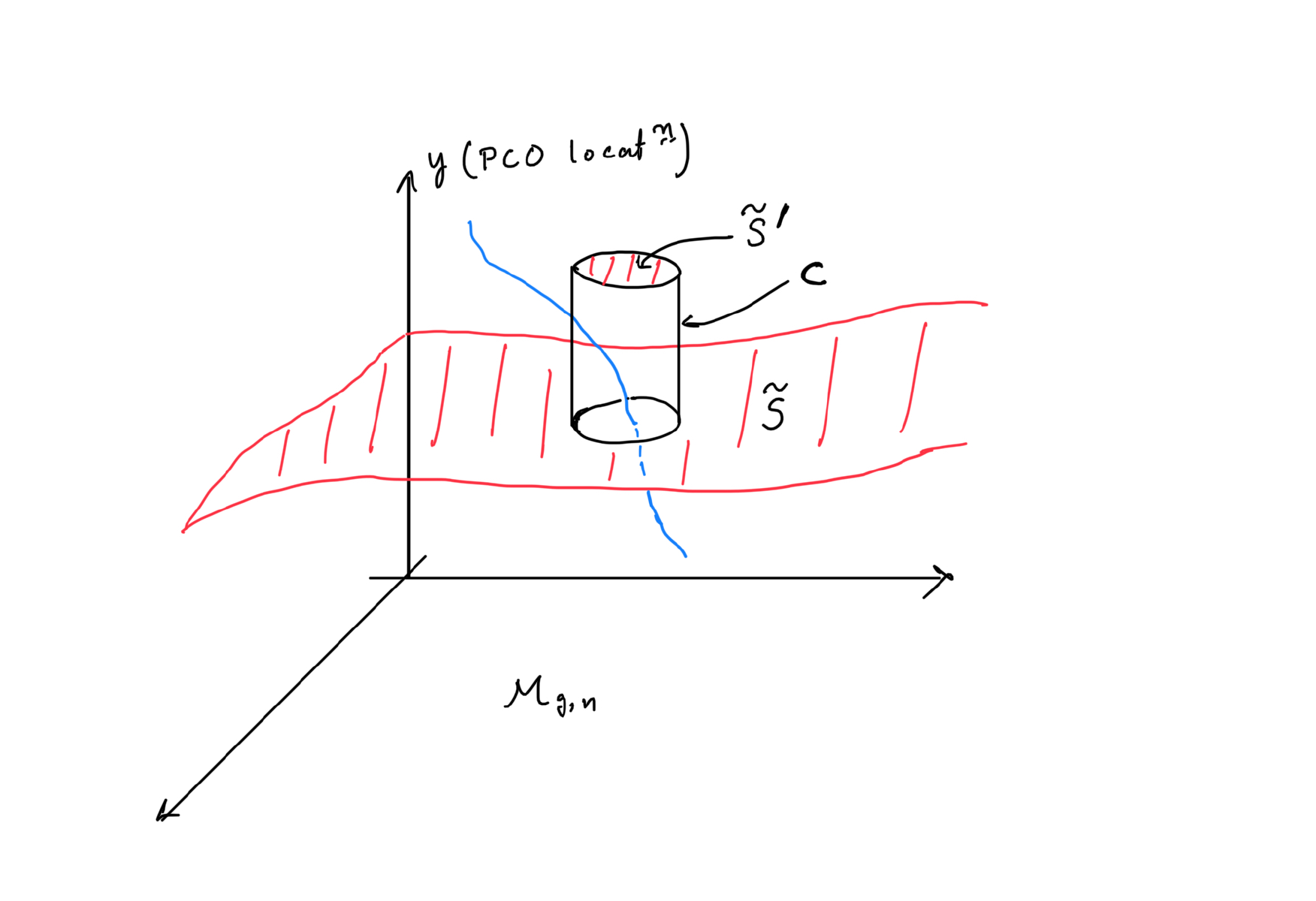"}
\caption{Vertical integration procedure to avoid spurious poles lying on a co-dimension 2 subspace (blue line).}
\label{fig:VInt}
\end{figure}

However there exists a more practical approach to avoid the spurious poles by moving the PCOs as described in \cite{Sen:2014pia}. It is this approach we are going to follow for our computation and here we summarise the main results. The starting point is to consider the PCO locations as fibre directions over the moduli space on top of the choice local coordinates and transition functions. So the choice of section now includes the PCO locations as well. Whenever we are near a spurious pole during the integration over a section, we move along the fibre directions specifying the PCO locations i.e. we move the PCO from its current location ($y_0$) to some other point ($y'$) keeping everything else fixed (see figure \ref{fig:VInt}).   
\\
So the final result is obtained by integrating over $\tilde{S}\cup C\cup \tilde{S}'$. Suppose now that we are moving a single PCO and $t$ parametrises the fibre direction of the PCO location then the integration over the cylinder $C$ yields,
\begin{equation}
\begin{aligned}
N_{\pi}\int\ \prod_{m=1}^{6g-6+2n-1}(\wedge d\tilde{t}_m)\wedge\int dt \big\langle\prod_{m=1}^{6g-6+2n-1} B_m\ B_{t}\prod_{\alpha}\chi(y_{\alpha})\ V_1\dots V_n\big\rangle,
\end{aligned}
\label{eq:Ex1}
\end{equation}   
where $N_{\pi}=(-2\pi i)^{-(3g-3+n)}$ and $B_t$ is given by,
$$B_{t}=\sum_{s}\oint_{C_s}\frac{\partial F_s}{\partial t}b(\sigma_s)d\sigma_s+\sum_{s}\oint_{C_s}\frac{\partial \bar{F}_s}{\partial t}\bar{b}(\bar{\sigma}_s)d\bar{\sigma}_s-\frac{1}{\chi(y)}\frac{\partial y}{\partial t}\partial\xi(y(t))=-\frac{1}{\chi(y)}\frac{\partial y}{\partial t}\partial\xi(y(t))$$
since none of the transition functions $F_s, \bar{F}_s$ depend on $t$. As result, using the fact that $\partial\xi$ is a total derivative of the field $\xi$ in the large Hilbert space we get,
\begin{equation}
N_{\pi}\int\ \prod_{m=1}^{6g-6+2n-1}(\wedge d\tilde{t}_m) \bigg\langle(\xi(y_0)-\xi(y'))\prod_{m=1}^{6g-6+2n-1} B_m\ \prod_{\alpha}\chi(y_{\alpha})\ V_1\dots V_n\bigg\rangle,
\end{equation}

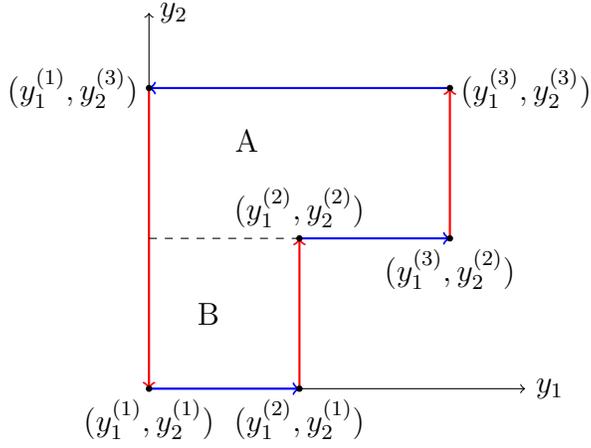
\begin{figure}[h!]
\centering
\begin{tikzpicture}
\draw[black, ->] (0,0) -- (5,0) node[anchor=west]{$y_1$};
\draw[black, ->] (0,0) -- (0,5) node[anchor=west]{$y_2$};
\draw[blue, ->, thick] (0,0) -- (2,0);
\draw[red, ->, thick] (2,0) -- (2,2);
\draw[blue, ->, thick] (2,2) -- (4,2);
\draw[red, ->, thick] (4,2) -- (4,4);
\draw[blue, ->, thick] (4,4) -- (0,4);
\draw[red, ->, thick] (0,4) -- (0,0);
\draw[black, dashed] (0,2) -- (2,2);
\filldraw[black] (1,3.3) circle (0pt) node[anchor=west]{A};
\filldraw[black] (0.5,1) circle (0pt) node[anchor=west]{B};
\filldraw[black] (0,0) circle (1pt) node[anchor=north]{$(y_1^{(1)},y_2^{(1)})$};
\filldraw[black] (2,0) circle (1pt) node[anchor=north]{$(y_1^{(2)},y_2^{(1)})$};
\filldraw[black] (2,2) circle (1pt) node[anchor=south]{$(y_1^{(2)},y_2^{(2)})$};
\filldraw[black] (4,2) circle (1pt) node[anchor=north]{$(y_1^{(3)},y_2^{(2)})$};
\filldraw[black] (4,4) circle (1pt) node[anchor=west]{$(y_1^{(3)},y_2^{(3)})$};
\filldraw[black] (0,4) circle (1pt) node[anchor=east]{$(y_1^{(1)},y_2^{(3)})$};
\end{tikzpicture}
\caption{Vertical hole over a point in the moduli space.}
\label{fig:hole}
\end{figure}

When we have to move more than one PCO, we move them one at a time but since changing the order in which we move them results in different paths (figure \ref{fig:hole}). As a result we have a hole over a point in the moduli space which we need to fill. We give here the result of moving two PCOs, for moving more number of PCOs the result easily generalises. 
Following the case of moving one PCO we now have for moving two PCOs,
\begin{equation}
\begin{aligned}
N_{\pi}\int\ \prod_{m=1}^{6g-6+2n-1}(\wedge d\tilde{t}_m)\wedge\int dt_1\wedge dt_2\big\langle\prod_{m=1}^{6g-6+2n-1} B_m\ B_{t_1}B_{t_2}\prod_{\alpha}\chi(y_{\alpha})\ V_1\dots V_n\big\rangle
\end{aligned}
\end{equation}   
where $t_1$ and $t_2$ parametrise the direction of the two PCO location $y_1$ and $y_2$, and $B_{t_1},B_{t_2}$ are given by,
$$B_{t_1}=-\frac{1}{\chi(y_1)}\frac{\partial y_1}{\partial t_1}\partial\xi(y_1(t_1)) \ ;\quad B_{t_2}=-\frac{1}{\chi(y_2)}\frac{\partial y_2}{\partial t_2}\partial\xi(y_2(t_2))\ .$$
The final result after filling this 2d vertical hole we have,
\begin{equation}
\begin{aligned}
&\int\ \prod_{m=1}^{6g-6+2n-1}(\wedge d\tilde{t}_m)\int_{A+B} dy_1\wedge dy_2\bigg\langle \partial \xi(y_1)\partial \xi(y_2)\dots\bigg\rangle =\\
&\hspace{1.5cm}\bigg\langle\bigg\lbrace\big(\xi(y_1^{(3)})-\xi(y_1^{(1)})\big)\big(\xi(y_2^{(3)})-\xi(y_2^{(2)})\big)+\big(\xi(y_1^{(2)})-\xi(y_1^{(1)})\big)\big(\xi(y_2^{(2)})-\xi(y_2^{(1)})\big)\bigg\rbrace\dots\bigg\rangle
\end{aligned}
\end{equation}

\section{NS 1 point function on genus 2 Riemann surface}
\label{sec:g2p1}
To carry out this computation,
\begin{itemize}
\item{we first make a choice of local coordinates on the 1 punctured Riemann surface of genus 2.}
\item{Then we write the amplitude with the Beltrami differentials and appropriate number of PCOs, and show that it is a total derivative on the Moduli space and hence gets contribution only from the boundary.}
\item{Finally we calculate the possible contributions from the boundary.}
\end{itemize}  
This technique of computing closely follows \cite{Atick:1987qy}.

\subsection{Choice of local coordinates}
\label{sec:main}
\begin{figure}[h!]
\centering
\includegraphics[scale=0.5]{"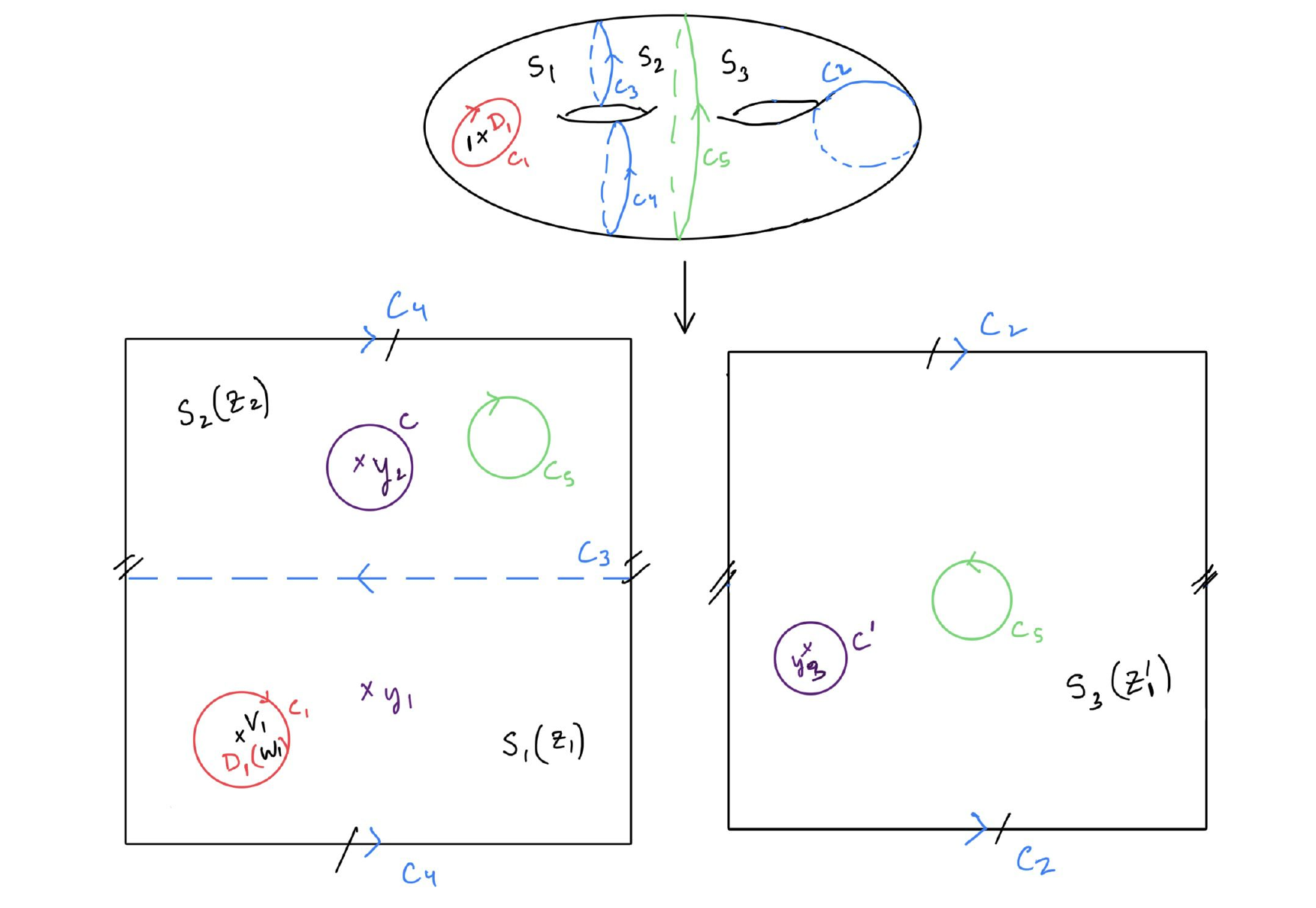"}
\caption{Pair of pants decomposition for genus 2 with 1 puncture}
\label{fig:2loop1pt}
\end{figure}

We know that the genus two Riemann surface with 1 NS punctures can be thought of as a union of $n=1$ Disk around the puncture, $2g-2+n=3$ spheres with 3 holes each, joined at $3g-3+2n=5$ circles (As in figure \ref{fig:2loop1pt}). We also have,
\\
\begin{eqnarray}
\text{Moduli space}\ \mathcal{M}_{2,1,0} &:=& 6g-6+2n=8\ \text{real i.e. 4 complex dimensions},\\
\text{Required no. of PCOs} &:=& 2g-2+n=3\ .
\end{eqnarray} 
We call the coordinate on $S_1$ as $z_1$, on $S_2$ as $z_2$ and on $S_3$ as $z'_1$. The coordinate on the disk $D_1$ around the first puncture is $w_1$ such that the location of the puncture in this coordinate is $w_1=0$. We depict these choice of coordinates on the bottom part in figure \ref{fig:2loop1pt}. 

With this choice of local coordinates we write down the transition functions as,
\begin{eqnarray}
z_1\ =\ w_1+\zeta_1 && \text{on $C_1$},\label{eq:cp1}\\
z_1\ =\ z_2 && \text{on $C_3$},\\
z_2\ =\ -\frac{q}{z'_1} && \text{on $C_5$},\\
z_1\ =\ z_2-\tau_1 && \text{on $C_4$},\\
z'_1\ =\ z'_1-\tau_2 && \text{on $C_2$}\ .\label{eq:cp2}
\end{eqnarray}
Here $\tau_1,\tau_2,\zeta_1,q$ are 4 complex parameters accounting for the 4 complex moduli that are required. Let us remark here once more that the fibre directions over the base moduli space also contain the locations of the PCOs. The violet points denote the locations of the 3 PCOs at $y_1,y_2$ and $y_3$. In our computation we will take $y_1\rightarrow p_1$ and $y_2\rightarrow p_2$ at the very end, for the reason mentioned in the introduction.
One can check easily that the above choice of local coordinates around the punctures along with the transition functions is manifestly gluing compatible. 

\subsection{The 1-point function as a total derivative using vertical integration procedure}
\label{sec:main1}
In accordance with \cite{Atick:1987qy} we do our computation in the large Hilbert space which means we include $\xi$ field at some arbitrary point $z_0$ to soak up the $\xi$ zero mode.

To get an invariant measure for integrating over this section we need the so called Beltrami differentials $B_m$ for each moduli variable $\tilde{t}_m$. For the superstring case they are defined as 
\begin{equation}
B_{m}=\sum_{s}\oint_{C_s}\frac{\partial F_s}{\partial \tilde{t}_m}b(\sigma_s)d\sigma_s+\sum_{s}\oint_{C_s}\frac{\partial \bar{F}_s}{\partial \tilde{t}_m}\bar{b}(\bar{\sigma}_s)d\bar{\sigma}_s-\frac{1}{\chi(y)}\frac{\partial y}{\partial \tilde{t}_m}\partial\xi(y(\tilde{t}_m)),
\end{equation} 
where $\sigma_s=F_s(\tau_s)$ defines the transition function between the coordinate patches $\sigma_s$ and $\tau_s$ on the circle $C_s$. For our choice of local coordinates and transition functions \eqref{eq:cp1}-\eqref{eq:cp2} the Beltrami differentials are given by (the PCO locations being independent of the the moduli parameters),
\begin{eqnarray}
B_{\tau_1}=-\oint_{C_4}b(z_1)dz_1, && B_{\bar{\tau}_1}=-\oint_{C_4}\bar{b}(\bar{z}_1)d\bar{z}_1\\
B_{\tau_2}=-\oint_{C_2}b(z'_1)dz'_1, && B_{\bar{\tau}_2}=-\oint_{C_2}\bar{b}(\bar{z}'_1)d\bar{z}'_1\\
B_{\zeta_1}=\oint_{C_1}b(w_1)dw_1, && B_{\bar{\zeta}_1}=\oint_{C_1}\bar{b}(\bar{w}_1)d\bar{w}_1\\
B_{q}=\oint_{C_5}q^{-1}z_2b(z_2)dz_2, && B_{\bar{q}}=\oint_{C_5}\bar{q}^{-1}\bar{z}_2\bar{b}(\bar{z}_2)d\bar{z}_2
\end{eqnarray}
Now we are ready to write the 2 loop 1 point amplitude $A^{(g=2)}_1$. Most of the treatment that we will follow in the rest of this subsection closely resembles \cite{Atick:1987qy}.

First we will write the general form of the amplitude for on-shell external states (i.e. the vertex, $V_1$, is a super-conformal primary of dimension (0,0) along with $L_0^-|V_i\rangle=b_0^-|V_i\rangle=0,\ \forall\ i=1,2$) in the NS sector (i.e. -1 picture number). 
\begin{equation}
V_1(z,\bar{z})=\varepsilon_{\mu}^a \bar{c}ce^{-\phi}\psi^{\mu}(z)\bar{J}_a(\bar{z})e^{ik.X(z,\bar{z})}\ ,\ \ \text{with $k^2=0$ and $\varepsilon_{\mu}^ak^{\mu}=0$.}
\end{equation} 
The 2 loop 1 point function is given by,
\begin{equation}
\begin{aligned}
A^{(2)}_1 =& (-2\pi i)^{-4}g_s^3\int d\tau_1\wedge d\bar{\tau}_1\wedge d\tau_2\wedge d\bar{\tau}_2\wedge d\zeta_1\wedge d\bar{\zeta}_1\wedge dq\wedge d\bar{q} \\
&\hspace{2cm}\bigg\langle B_{\tau_1}B_{\bar{\tau}_1}B_{\tau_2}B_{\bar{\tau}_2}B_{\zeta_1}B_{\bar{\zeta}_1}B_{q}B_{\bar{q}}\ \xi(z_0)\prod_{\alpha=1}^{3}\chi(y_{\alpha})f_1\circ V_1(0)\bigg\rangle
\end{aligned}
\label{eq:A2l1}
\end{equation}
We first bring the PCO at $y_3$ on top of $V_1$ to make it a 0-picture vertex,
\begin{equation}
\begin{aligned}
&\ e^{\phi}T_F(z)V_1(w,\bar{w})\sim i\sqrt{2}\varepsilon_{\mu}^a\bar{c}c(z-w)\left(\frac{\partial X^{\mu}(w)}{2(z-w)}-\frac{ik.\psi}{2(z-w)}\psi^{\mu}(w)\right)\bar{J}_a(\bar{w})e^{ik.X(w,\bar{w})}\\
&\Rightarrow V_1^{(0)}(w,\bar{w})=\ \frac{i}{\sqrt{2}}\varepsilon_{\mu}^a\bar{c}c\left(\partial X^{\mu}(w)-i(k.\psi)\psi^{\mu}(w)\right)\bar{J}_a(\bar{w})e^{ik.X(w,\bar{w})}\ .
\end{aligned}
\end{equation}
Finally we write the 0-picture vertex as spacetime super current acting on a 1/2-picture vertex as in \cite{Atick:1987qy}. We have,
\begin{equation}
\begin{aligned}
&\langle V_1^{(0)}(k,\zeta_1,\bar{\zeta}_1)\rangle=\langle\oint_{\zeta_1} \frac{dx}{2\pi i} \mathcal{J}(x)V_1^{(\frac{1}{2})}(k,\zeta_1,\bar{\zeta}_1)\rangle\\
&\text{with 10d super current\ }\mathcal{J}(x)=e^{-\frac{\phi}{2}}u^{\beta}(0)S_{\beta}(x)\text{ and}\ ,\\
&V_1^{(\frac{1}{2})}(k,\zeta_1,\bar{\zeta}_1)=\epsilon_a v^{\alpha}(k)\bar{c}c\left\lbrace e^{\frac{\phi}{2}}\left[\partial X^{\mu}-\frac{1}{4}i(k.\psi)\psi^{\mu}\right](\gamma_{\mu})_{\alpha\alpha'}S^{\alpha'}+\frac{1}{2}e^{3\phi/2}\eta bS_{\alpha}\right\rbrace(\zeta_1)\bar{J}^a(\bar{\zeta}_1)e^{ik.X}.
\end{aligned}
\end{equation}
 Notice that the super current has no anti-holomorphic piece since we are working with heterotic strings. Also the term containing $e^{3\phi/2}$ in $V_1^{(\frac{1}{2})}$ does not contribute to the amplitude due to $\phi$ charge conservation. The above result is illustrated in \cite{Friedan:1985ge}\footnote{In our case we have set the momentum of the -1/2 picture vertex to 0 to get the 10d spacetime super current.}. So finally we have the integrand in \eqref{eq:A2l1} given by,
 \begin{equation}
 \bigg\langle B_{\tau_1}B_{\bar{\tau}_1}B_{\tau_2}B_{\bar{\tau}_2}B_{\zeta_1}B_{\bar{\zeta}_1}B_{q}B_{\bar{q}}\ \xi(z_0)\prod_{\alpha=1}^{2}\chi(y_{\alpha})\oint_{\zeta_1} \frac{dx}{2\pi i} \mathcal{J}(x)V_1^{(\frac{1}{2})}(k,\zeta_1,\bar{\zeta}_1)\bigg\rangle
 \label{eq:A2l1_2}
 \end{equation}
 The $x$ contour can now be pulled out and closed to some point away from $\zeta_1$ since the surface is a closed one. In this process we pickup the residues from different spurious poles in the $x$ plane whose location ($r_l$ say) are a function of $y_1,y_2$ and $\zeta_1$. Now if the PCO at $y_1$ is moved to some other point $\tilde{y}_1$ such that the new pole locations are away from $r_l$ then the contour integration simply vanishes. So we can rewrite \eqref{eq:A2l1_2} using $\chi(y_1)-\chi(\tilde{y}_1)=\lbrace Q_B,\xi(y_1)-\xi(\tilde{y}_1)\rbrace$ as,
 \begin{equation}
 \sum_{r_l}\oint_{r_l} dx\bigg\langle B_{\tau_1}B_{\bar{\tau}_1}B_{\tau_2}B_{\bar{\tau}_2}B_{\zeta_1}B_{\bar{\zeta}_1}B_{q}B_{\bar{q}}\ \xi(z_0)Q_B(\xi(y_1)-\xi(\tilde{y}_1))\chi(y_2) \mathcal{J}(x)V_1^{(\frac{1}{2})}(k,\zeta_1,\bar{\zeta}_1)\bigg\rangle
 \label{eq:A2l1_3}
 \end{equation}

We now use the following well known identity to move the BRST operator on the other operators.
\begin{equation}
\begin{aligned}
&\Omega_p^{(g,n)}(Q_BV_1,V_2,\dots,V_n)+(-1)^{V_1}\Omega_p^{(g,n)}(V_1,Q_BV_2,\dots,V_n)+\dots\\
&\dots+(-1)^{V_1+V_2+...+V_{n-1}}\Omega_p^{(g,n)}(V_1,V_2,\dots,Q_BV_n)=(-1)^pd\Omega_{p-1}^{(g,n)}(V_1,V_2,\dots,V_n)\ .
\end{aligned}
\label{eq:main_id}
\end{equation}
This identity uses the fact that $\lbrace Q_B,b(z)\rbrace=T(z)$.
Here $\Omega_p^{g,n}$ describes a $p$-form on a section of the fibre directions over the moduli space of genus $g$ and $n$-punctures. In our case we have, as mentioned $\Omega_8^{(2,1)}(\chi_1,\chi_2,\chi_3,V_1)$, but note the following points.
\begin{itemize}
\item{Since we are working with on-shell external states we have by definition $Q_BV_1=0$.}
\item{Also for the PCO's we have $Q_B\chi=0$ since we know that $Q_B^2=0$.}
\item{Since $(\xi(y_1)-\xi(\tilde{y}_1))$ is in the small Hilbert space the term $\langle\dots Q_B\xi(z_0)(\xi(y_1)-\xi(\tilde{y}_1))\chi(y_2)\dots\rangle$ vanishes due to the fact that there is no field to soak up the $\xi$ zero mode. }
\end{itemize}

As result our integrand turns out to be a total derivative on the moduli space. Following \cite{Atick:1987qy} we know that the boundary is a real co-dimension 2 surface ($q=0$) and hence the boundary contribution vanishes unless the integrand becomes singular. Near the boundary $q\rightarrow 0$ we have,
\begin{equation}
\begin{aligned}
\lim_{a\rightarrow 0}\int_{|q|\geq a} dq\wedge d\bar{q}\ \frac{\partial}{\partial q}(F(q,\bar{q}))=\lim_{a\rightarrow 0}\int_0^{2\pi}d\theta\ (\bar{q}F(q,\bar{q}))|_{r=a}\ .
\end{aligned}
\label{eq:bdryterm}
\end{equation} 
Here we have used the parametrization $$q=re^{i\theta},\quad\Rightarrow\quad \bar{q}=re^{-i\theta}\ ,$$ and the only assumption is that $F(q,\bar{q})$ is analytic in $\theta$. From \eqref{eq:bdryterm} we see that as $a\rightarrow 0$, non zero contribution comes from the $1/\bar{q}$ divergence piece of $F(q,\bar{q})$ in the $q\rightarrow 0$ limit. 

For the current case at hand we put $z_0=y_1$ to get,
\begin{equation}
\begin{aligned}
F(q,\bar{q}) =& (-2\pi i)^{-4}g_s^3\int d\tau_1\wedge d\bar{\tau}_1\wedge d\tau_2\wedge d\bar{\tau}_2\wedge d\zeta_1\wedge d\bar{\zeta}_1 \\
&\sum_{r_l}\oint_{r_l} \frac{dx}{2\pi i}\bigg\langle B_{\tau_1}B_{\bar{\tau}_1}B_{\tau_2}B_{\bar{\tau}_2}B_{\zeta_1}B_{\bar{\zeta}_1}B_{\bar{q}}\ \xi(\tilde{y}_1)\xi(y_1)\chi(y_2)\mathcal{J}(x)V_1^{(\frac{1}{2})}(k,\zeta_1,\bar{\zeta}_1)\bigg\rangle
\end{aligned}
\label{eq:A2l1_3}
\end{equation}

\subsection{Contribution due to the $\xi,\ \eta,\ \phi$ system}
\label{sec:1pt_sghost}
In \eqref{eq:A2l1_3} most contributions vanish due to $\phi$ charge conservation (for genus 2 surface) i.e. $\sum q\neq 2$. The only non vanishing contribution is given by,
\begin{eqnarray}
F(q,\bar{q})&=&-\frac{1}{4}(-2\pi i)^{-4}g_s^3\lim_{y_2\rightarrow y'_2}\int d\tau_1\wedge d\bar{\tau}_1\wedge d\tau_2\wedge d\bar{\tau}_2\wedge d\zeta_1\wedge d\bar{\zeta}_1 \left(2\frac{\partial}{\partial y_2}+\frac{\partial}{\partial y'_2}\right)G(y_2,y'_2)\nonumber \\
\\
\text{with,}\nonumber\\
G(y_2,y'_2)&=&\sum_{r_l}\oint_{r_l} \frac{dx}{2\pi i}\bigg\langle B_{\tau_1}B_{\bar{\tau}_1}B_{\tau_2}B_{\bar{\tau}_2}B_{\zeta_1}B_{\bar{\zeta}_1}B_{\bar{q}}\ \xi(\tilde{y}_1)\xi(y_1)\eta(y_2)e^{2\phi(y'_2)}b(y'_2)\mathcal{J}(x)V_1^{(\frac{1}{2})}(k,\zeta_1,\bar{\zeta}_1)\bigg\rangle\nonumber\\
\end{eqnarray}
The superghost part of the correlation function is given by,
\begin{equation}
\begin{aligned}
&\left\langle \xi(\tilde{y}_1)\xi(y_1)\eta(y_2)e^{2\phi(y'_2)}e^{-\frac{\phi(x)}{2}}e^{\frac{\phi(\zeta_1)}{2}}\right\rangle \\ 
&\ =\ (Z_1)^{\frac{1}{2}}\frac{\vartheta[\delta]\left(\vec{\tilde{y}}_1+\vec{y}_1-2\vec{y}_2+2\vec{y}^{\ \prime}_2-\frac{\vec{x}}{2}+\frac{\vec{\zeta}_1}{2}-2\vec{\Delta}\right)}{\vartheta[\delta]\left(\vec{y}_1-\vec{y}_2+2\vec{y}^{\ \prime}_2-\frac{\vec{x}}{2}+\frac{\vec{\zeta}_1}{2}-2\vec{\Delta}\right)\vartheta[\delta]\left(\vec{\tilde{y}}_1-\vec{y}_2+2\vec{y}^{\ \prime}_2-\frac{\vec{x}}{2}+\frac{\vec{\zeta}_1}{2}-2\vec{\Delta}\right)}\\
&\ \ \ \frac{E(\tilde{y}_1,y_1)E(x,y'_2)(E(x,\zeta_1))^{\frac{1}{4}}}{E(\zeta_1,y'_2)E(\tilde{y}_1,y_2)E(y_1,y_2)}\frac{\sigma(x)}{\sigma(\zeta_1)(\sigma(y'_2))^4}\ .
\end{aligned}
\label{eq:superghost}
\end{equation}
Here we have used \eqref{eq:largeHilbert}. Now we have to take the limit $q\rightarrow 0$ to see what is the contribution at the boundary. The results that we will be using exhaustively from here on are given in the appendix of \cite{Atick:1987qy} which we present in the appendix \ref{app:1} for the sake of completeness. 

Let us first note that the spurious poles of the above function comes from the zero of $\vartheta[\delta]\left(\vec{y}_1-\vec{y}_2+2\vec{y}^{\ \prime}_2-\frac{\vec{x}}{2}+\frac{\vec{\zeta}_1}{2}-2\vec{\Delta}\right)$ in the $x$ plane. There are a total of $2^{2g-2}g=8$ spurious poles for genus $g=2$ surface. In the limit $q\rightarrow 0$ we take $y_1$ and $\zeta_1$ on the torus $T_1$ with modular parameter $\tau_1$ and $\tilde{y}_1,y_2$ and $y'_2$ are on the other torus $T_2$ with modular parameter $\tau_2$. Of course $\zeta_1$ is integrated over so we need to also consider the case when $\zeta_1$ lies on $T_2$ but that we will look into later, for now keep $\zeta_1$ on $T_1$. Now let us see what happens when,
\begin{enumerate}
\item{$x$ lies on $T_1$: In this case we get for $q\rightarrow 0$,
\begin{equation}
\begin{aligned}
&\vartheta[\delta]\left(\vec{y}_1-\vec{y}_2+2\vec{y}^{\ \prime}_2-\frac{\vec{x}}{2}+\frac{\vec{\zeta}_1}{2}-2\vec{\Delta}\right)\\
&\rightarrow
\vartheta\begin{bmatrix}
a_1\\
b_1
\end{bmatrix}\left(y_1-\frac{x}{2}+\frac{\zeta_1}{2}-p_1|\tau_1\right)
\vartheta\begin{bmatrix}
a_2\\
b_2
\end{bmatrix}\left(2y'_2-y_2-p_2|\tau_2\right)\ .
\end{aligned}
\end{equation}
For a given spin structure $(a_1,b_1)$ the theta function vanishes at\footnote{Note that for $T_1$, $\hat{a}_1,\hat{b}_1 \in \lbrace 0,\frac{1}{2}\rbrace$ s.t. $\vartheta\begin{bmatrix}
a_1\\
b_1
\end{bmatrix}\left(z-\hat{a}_1\tau_1-\hat{b}_1|\tau_1\right)\sim \vartheta\begin{bmatrix}
1/2\\
1/2
\end{bmatrix}\left(z|\tau_1\right)
$ upto some phase factor. Similarly for $T_2$, $\hat{a}_2,\hat{b}_2 \in \lbrace 0,\frac{1}{2}\rbrace$.}, $$-\frac{x}{2}=p_1-y_1-\frac{\zeta_1}{2}+\hat{a}_1\tau_1+\hat{b}_1$$ with four different spin structures $(a_2,b_2)$, which accounts for 4 out of 8 spurious poles. One can check that in \eqref{eq:superghost} the residue at these poles contain,
\begin{equation}
\begin{aligned}
&\vartheta[\delta]\left(\vec{\tilde{y}}_1+\vec{y}_1-2\vec{y}_2+2\vec{y}^{\ \prime}_2-\frac{\vec{x}}{2}+\frac{\vec{\zeta}_1}{2}-2\vec{\Delta}\right)\\
&\rightarrow
\vartheta\begin{bmatrix}
a_1\\
b_1
\end{bmatrix}\left(y_1-\frac{x}{2}+\frac{\zeta_1}{2}-p_1|\tau_1\right)
\vartheta\begin{bmatrix}
a_2\\
b_2
\end{bmatrix}\left(\vec{\tilde{y}}_1+2y'_2-2y_2-p_2|\tau_2\right)\ ,
\end{aligned}
\end{equation}
which vanishes exactly at the points where the poles are situated. So contribution from these 4 poles lying on $T_1$ simply vanish.}
\item{$x$ lies on $T_2$: In the $q\rightarrow 0$ limit we get,
\begin{equation}
\begin{aligned}
&\vartheta[\delta]\left(\vec{y}_1-\vec{y}_2+2\vec{y}^{\ \prime}_2-\frac{\vec{x}}{2}+\frac{\vec{\zeta}_1}{2}-2\vec{\Delta}\right)\\
&\rightarrow
\vartheta\begin{bmatrix}
a_1\\
b_1
\end{bmatrix}\left(y_1+\frac{\zeta_1}{2}-\frac{3p_1}{2}|\tau_1\right)
\vartheta\begin{bmatrix}
a_2\\
b_2
\end{bmatrix}\left(2y'_2-y_2-\frac{x}{2}-\frac{p_2}{2}|\tau_2\right)\ .
\end{aligned}
\end{equation}
This time for a specific $(a_2,b_2)$ the above function vanishes at $$\frac{x}{2}=2y'_2-y_2-\frac{p_2}{2}+\hat{a}_2\tau_2+\hat{b}_2$$ with four different spin structures $(a_1,b_1)$ which accounts for the left over 4 poles. Below, we just write down the final result for the residues at these poles. This can be easily checked using appendix \ref{app:1}. 
\begin{equation}
\begin{aligned}
&\text{In the limit $q\rightarrow 0$ we get,}\\
&q^{3/8}\eta_d(\tau_1)\eta_d(\tau_2)\frac{\vartheta_1(3y'_2-2y_2-p_2|\tau_2)\left[\vartheta_1(4y'_2-2y_2-2p_2|\tau_2)\vartheta_1(\zeta_1-p_1|\tau_1)\right]^{1/4}}{\vartheta_{a_1b_1}\left(\frac{\zeta_1}{2}-\frac{p_1}{2}|\tau_1\right)\vartheta_{1}\left(4y'_2-2y_2-2p_2|\tau_2\right)\left[\vartheta'_1(0|\tau_1)\right]^{1/4}\left[\vartheta'_1(0|\tau_2)\right]^{9/4}},
\end{aligned}
\end{equation}
upto some overall phase. Here $\eta_d$ denotes the Dedekind eta function and we have introduced the following notations for the sake of brevity,
$$\vartheta_{ab}(z|\tau)\equiv\vartheta\begin{bmatrix}
a\\
b
\end{bmatrix}(z|\tau)\ ,\quad \vartheta_{1}(z|\tau)\equiv\vartheta\begin{bmatrix}
1/2\\
1/2
\end{bmatrix}(z|\tau)\ ,\quad $$
$$\Rightarrow\quad\vartheta_1(0|\tau)=0\ , \quad \vartheta'_1(0|\tau)=\left(\frac{\partial }{\partial z}\vartheta_1(z|\tau)\right)\bigg|_{z=0}$$}
\end{enumerate}

\subsection{Contribution due to $b,c$ ghost and the matter sector}
\label{sec:1pt_bcmatter}
For this case we need to first note that $B_{\zeta_1}B_{\bar{\zeta}_1}$ acts on the $\bar{c}(\bar{\zeta}_1)c(\zeta_1)$ to give $1/|w-\zeta_1|^2$ so that the contour integral on $w$ and $\bar{w}$ gives identity and we are left with the following $b,c$ correlation function,
\begin{equation}
\langle B_{\tau_1}B_{\bar{\tau}_1}B_{\tau_2}B_{\bar{\tau}_2}B_{\bar{q}}\ b(y'_2)\rangle=\frac{\bar{q}^{-2}}{q}|\eta_d(\tau_1)|^4|\eta_d(\tau_2)|^4\left(\frac{\vartheta'_1(0|\tau_2)}{\vartheta_1(y'_2-p_2|\tau_2)}\right)^2\ .
\end{equation}
Thus, so far we have,
\begin{equation}
\begin{aligned}
G(y_2,y'_2)=&q^{-5/8}\bar{q}^{-2}\eta_d(\tau_1)^3\eta_d(\tau_2)^3\bar{\eta}_d(\bar{\tau}_1)^2\bar{\eta}_d(\bar{\tau}_2)^2\left(\frac{\vartheta'_1(0|\tau_2)}{\vartheta_1(y'_2-p_2|\tau_2)}\right)^2\bigg\langle\text{matter}\bigg\rangle\\
&\frac{\vartheta_1(3y'_2-2y_2-p_2|\tau_2)\left[\vartheta_1(4y'_2-2y_2-2p_2|\tau_2)\vartheta_1(\zeta_1-p_1|\tau_1)\right]^{1/4}}{\vartheta_{a_1b_1}\left(\frac{\zeta_1}{2}-\frac{p_1}{2}|\tau_1\right)\vartheta_{1}\left(4y'_2-2y_2-2p_2|\tau_2\right)\left[\vartheta'_1(0|\tau_1)\right]^{1/4}\left[\vartheta'_1(0|\tau_2)\right]^{9/4}}\ .
\end{aligned}
\label{eq:G1}
\end{equation}

The above expression implies that we need only compute the contribution from the matter sector which is proportional to $q^{5/8}\bar{q}$ so that we get the $1/\bar{q}$ piece of $F(q,\bar{q})$. The matter part of the correlation function is,
\begin{equation}
u^{\alpha}v^{\alpha'}(\gamma_{\mu})_{\beta\alpha'}\epsilon_a\left\langle S_{\alpha}(x)S^{\beta}(\zeta_1)\left(\partial X^{\mu}-\frac{1}{4}i(k.\psi)\psi^{\mu}\right)\bar{J}^a(\bar{\zeta}_1)e^{ik.X}\right\rangle\sim(2\pi)^{10}\delta^{10}(k)(\bar{u}\gamma_{\mu}v)\epsilon_a\langle\dots\rangle.
\end{equation}
Due to the overall momentum conserving delta function we can put $k=0$ and drop the term $(k.\psi)\psi^{\mu}$.
For definiteness we take, $\langle S_{\alpha}(x)S^{\beta}(\zeta_1)\dots\rangle\sim \delta_{\alpha}^{\ \beta}\langle S_{1+}..S_{5+}(x)S_1^{\ -}..S_5^{\ -}(\zeta_1)\dots\rangle$. Now from the factorization theorem \cite{Friedan:1986ua}, we know that
\begin{equation}
\langle A_1(z_1)A_2(z_2)\rangle \sim \sum_{\varphi}\langle A_1(z_1)\varphi(p_1)\rangle_{T_1}\langle \varphi^{\dagger}(p_2)A_2(z_2)\rangle_{T_2}q^{h_{\varphi}}\bar{q}^{\bar{h}_{\varphi}}\ .
\label{eq:factorization}
\end{equation}
In the current scenario we have $A_1(z_1)\equiv S_1^{\ -}(\zeta_1)S_2^{\ -}(\zeta_1)S_3^{\ -}(\zeta_1)S_4^{\ -}(\zeta_1)S_5^{\ -}(\zeta_1)$ and $A_2(z_2)\equiv S_{1+}(x)S_{2+}(x)S_{3+}(x)S_{4+}(x)S_{5+}(x)$.
 
The operator with conformal dimension $\left(1,\frac{5}{8}\right)$ that contributes to the sum on the r.h.s above is,
$$\varphi(p_1)\equiv S_{1+}(p_1)S_{2+}(p_1)S_{3+}(p_1)S_{4+}(p_1)S_{5+}(p_1)\bar{J}^b(p_1)$$
$$\Rightarrow\varphi^{\dagger}(p_2)\equiv S_1^{\ -}(p_2)S_2^{\ -}(p_2)S_3^{\ -}(p_2)S_4^{\ -}(p_2)S_5^{\ -}(p_2)\bar{J}^b(p_2)$$
For uncompactified theories this is the only possibility although for compactified theories one may construct other type of operators, for more details see \cite{Atick:1987qy}. 

So the relevant contribution to matter correlation function is given by\footnote{The suffix on the correlator denotes on which torus we should compute each part of the correlation function.},
\begin{equation}
\begin{aligned}
&(\bar{u}\gamma_{\mu}v)\epsilon_aq^{5/8}\bar{q}\langle \partial X^{\mu}(\zeta_1)\bar{J}^a(\bar{\zeta}_1)\rangle_{T_1} \\
&\langle S_1^{\ -}(p_2)S_2^{\ -}(p_2)S_3^{\ -}(p_2)S_4^{\ -}(p_2)S_5^{\ -}(p_2)\bar{J}^b(p_2)S_{1+}(x)S_{2+}(x)S_{3+}(x)S_{4+}(x)S_{5+}(x)\rangle_{a_2b_2}\\
&\langle S_{1+}(p_1)S_{2+}(p_1)S_{3+}(p_1)S_{4+}(p_1)S_{5+}(p_1)\bar{J}^b(p_1)S_1^{\ -}(\zeta_1)S_2^{\ -}(\zeta_1)S_3^{\ -}(\zeta_1)S_4^{\ -}(\zeta_1)S_5^{\ -}(\zeta_1)\rangle_{a_1b_1}\ .
\end{aligned}
\end{equation}
The result of this correlation function in our case i.e. the uncompactified theory is,
\begin{equation}
\begin{aligned}
&(\bar{u}\gamma_{\mu}v)\epsilon_aq^{5/8}\bar{q}\langle \partial X^{\mu}(\zeta_1)\bar{J}^a(\bar{\zeta}_1)\rangle\\
&\tilde{K}(\tau_1)\tilde{K}(\tau_2)(\eta_d(\tau_1)\eta_d(\tau_2))^{-3}(\bar{\eta}_d(\bar{\tau}_1)\bar{\eta}_d(\bar{\tau}_2))^{-2}\varepsilon(a_1,b_1)\varepsilon(a_2,b_2)\left[\vartheta_{a_1b_1}\left(\frac{\zeta_1}{2}-\frac{p_1}{2}|\tau_1\right)\right]^5\\
&\left[\vartheta_{a_2b_2}\left(\frac{p_2}{2}-\frac{x}{2}|\tau_2\right)\right]^5\left[\vartheta_{1}\left(\zeta_1-p_1|\tau_1\right)\right]^{-\frac{5}{4}}\left[\vartheta_{1}\left(p_2-x|\tau_1\right)\right]^{-\frac{5}{4}}\left[\vartheta'_{1}\left(0|\tau_1\right)\vartheta'_{1}\left(0|\tau_2\right)\right]^{\frac{5}{4}}
\end{aligned}
\label{eq:matter1}
\end{equation} 
Notice that we need to sum over the spin structure $(a_1,b_1)$ of $T_1$\footnote{$\varepsilon(a_1,b_1)=1$ for $(a_1,b_1)=(1/2,1/2), (0,0)$ and $\varepsilon(a_1,b_1)=-1$ for $(a_1,b_1)=(1/2,0), (0,1/2)$.} whereas as for $T_2$ the spin structure is a fixed one for which the poles are situated at, $$\frac{x}{2}=2y'_2-y_2-\frac{p_2}{2}+\hat{a}_2\tau_2+\hat{b}_2$$ and we are computing the residues at these poles.
Putting \eqref{eq:G1} and \eqref{eq:matter1} together and carrying out the sum over $(a_1,b_1)$ we get,
\begin{equation}
\sum_{a_1,b_1}\varepsilon(a_1,b_1)\left[\vartheta_{a_1b_1}\left(\frac{\zeta_1}{2}-\frac{p_1}{2}|\tau_1\right)\right]^4\times(\dots)\ =\ 0,
\end{equation}
where the $(\dots)$ denote all the other pieces and they are independent of $(a_1,b_1)$. This sum vanishes due to the Riemann theta function identity,
\begin{eqnarray}
\sum_{\delta}\varepsilon(\delta)\vartheta[\delta](z_1)\vartheta[\delta](z_2)\vartheta[\delta](z_3)\vartheta[\delta](z_4)&=&2(\vartheta_1(z_1+z_2+z_3+z_4)\vartheta_1(z_1+z_2-z_3-z_4)\nonumber\\
&&\vartheta_1(z_1-z_2-z_3+z_4)\vartheta_1(z_1-z_2+z_3-z_4))
\label{eq:theta_id}
\end{eqnarray}

So we see that the integrand vanishes in the region of integration when $\zeta_1$ lies on $T_1$. Let us now consider the remaining region of integration where $\zeta_1$ lies on $T_2$. One can easily see that when $x$ lies on $T_1$, the residues at the 4 poles lying on $T_1$ simply vanishes. For a given spin structure $(a_1,b_1)$ location of these poles are given by,
$$\frac{x}{2}=y_1-\frac{p_1}{2}+\hat{a}_1\tau_1+\hat{b}_1\ ,\quad\text{for 4 different spin structures $(a_2,b_2)$.}$$
Thus let us consider the case when $x$ lies on $T_2$. For a given spin structure $(a_2,b_2)$ the location of 4 poles on $T_2$ are given by, $$\frac{x}{2}=2y'_2-y_2+\frac{\zeta_1}{2}-p_2+\hat{a}_2\tau_2+\hat{b}_2\ ,\quad\text{for 4 different spin structures $(a_1,b_1)$.}$$
We can now follow the same steps described previously in this section to get the full contribution from the residues at the poles on $T_2$. Notice that the contribution from the superghost and the conformal $b,c$ ghost part goes as,
\begin{equation}
\sim q^{-1}\bar{q}^{-2}\frac{1}{\vartheta_{a_1b_1}(0|\tau_1)}\times(\dots)\ ,
\end{equation}
where the $(\dots)$ denotes the total contribution independent of $q,\bar{q}$ and $(a_1,b_1)$. Finally, we turn to the matter contribution and note that the factorization theorem \eqref{eq:factorization} now applies to the case $A_2(z_2)\equiv S_1^{\ -}(\zeta_1)S_2^{\ -}(\zeta_1)S_3^{\ -}(\zeta_1)S_4^{\ -}(\zeta_1)S_5^{\ -}(\zeta_1)S_{1+}(x)S_{2+}(x)S_{3+}(x)S_{4+}(x)S_{5+}(x)$ and $A_1(z_1)\equiv 1$.
Also notice that now the relevant piece must contain the operator with conformal dimension $(1,1)$ so that we get the $1/\bar{q}$ contribution to $F(q,\bar{q})$. Although there are more than one such operator present in the theory, the relevant one is\footnote{The operator of the form $\varphi=\partial XJ^b$ does not contribute since $k=0$ for the one point function and hence $\delta^{10}(k)\langle \partial X(p_2) e^{ik.X(\zeta_1,\bar{\zeta}_1)}\rangle=0$}, $$\varphi(p_1)=\bar{\Psi}_i(p_1)\Psi_i(p_1)\bar{J}^b(p_1)$$$$\varphi^{\dagger}(p_2)=\bar{\Psi}_i(p_2)\Psi_i(p_2)\bar{J}^b(p_2)$$
where $i=1,..,5$ and $\Psi$ denotes a world-sheet complex fermion built from a pair of Majorana fermions. As a result the full correlation function (ghost and matter) on $T_1$ from all the holomorphic fermions is simply, $$\frac{1}{\bar{q}}\sum_{a_1,b_1}\varepsilon(a_1,b_1)\vartheta_{a_1b_1}(0|\tau_1)^4\times(\dots)=0\ \text{using \eqref{eq:theta_id}}.$$
Again $(\dots)$ being the contribution independent of $(a_1,b_1)$.
This implies that,
\begin{equation}
G(y_2,y'_2)=0\ ,\quad\Rightarrow\quad F(q,\bar{q})=0\ ,\quad\Rightarrow\quad A_1^{(2)}=0
\end{equation}
identically and thus they remain zero even as we take $y_1\rightarrow p_1$ and $y_2\rightarrow p_2$.
Hence, the 1-point function of the massless field vanishes at two loop order as expected.

\section{Renormalised mass at 2 loop order}
\label{sec:g2p2}
In the previous section we computed the full 1 point function but now we want to compute only a part of the 2 point function that constitutes the 1PI subspace of the amplitude. The basic procedure follows the previous section closely i.e. first we write the full amplitude as a total derivative on the moduli space using the vertical integration procedure. This implies that the contribution comes from the boundary and we figure out this contribution from different regions of integration of the vertex location separately. It is in this second step where we have to carefully identify and drop the regions of integration that constitute the 1PR subspace. This will leave us with the desired 1PI subspace and our result will give the renormalised mass. As will be clear from the treatment below that the full ghost contribution to the renormalised mass is independent of whether we consider the massless or massive vertex operators.

\subsection{Choice of local coordinates}
\label{sec:g2p2_1}
\begin{figure}[h!]
\centering
\includegraphics[scale=0.5]{"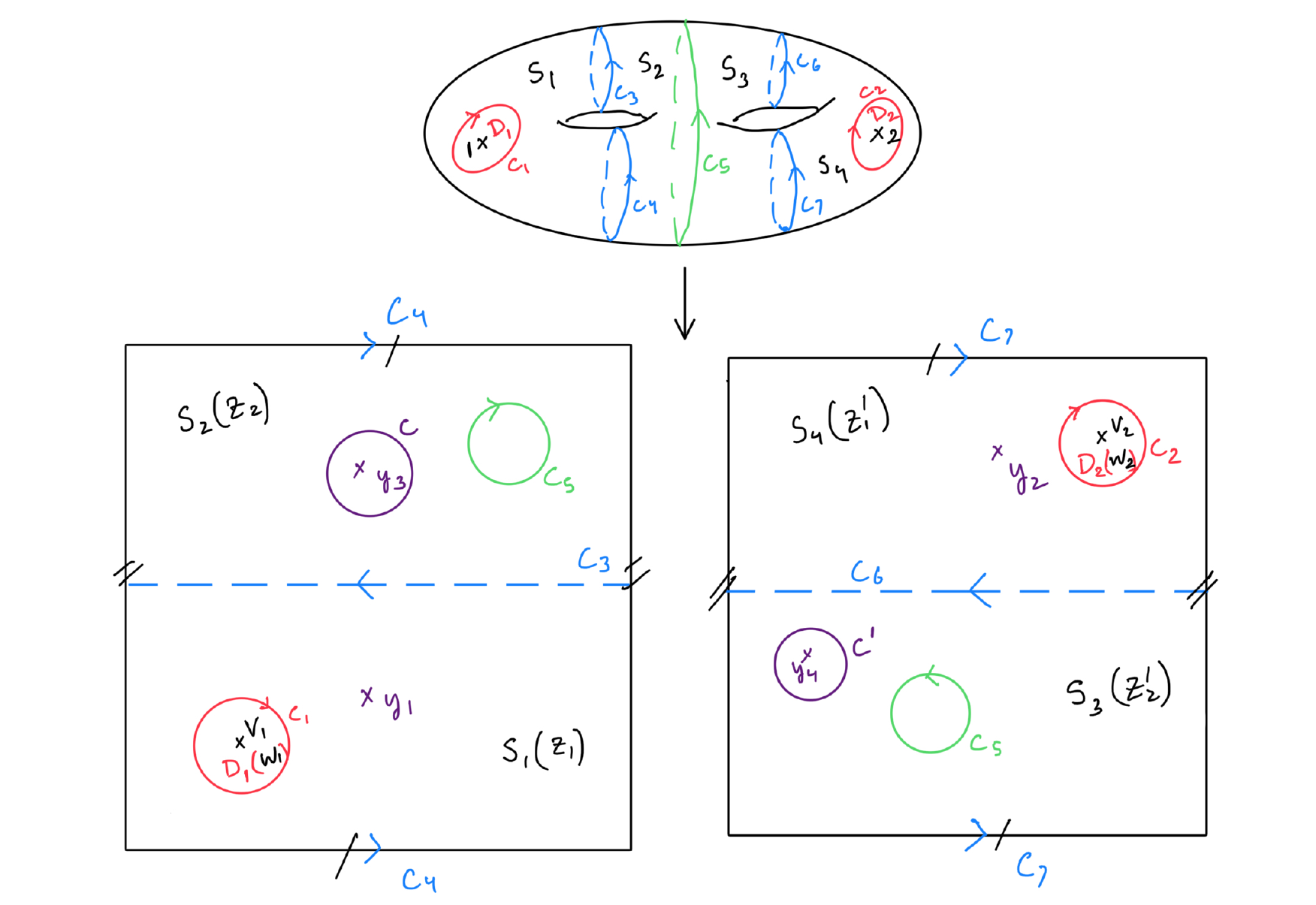"}
\caption{Pair of pants decomposition for genus 2 with 1 puncture}
\label{fig:2loop2pt}
\end{figure}

We now have a genus two Riemann surface with 2 NS punctures which is a union of $n=2$ Disks, $2g-2+n=4$ spheres with three holes each, joined at $3g-3+2n=7$ circles (figure \ref{fig:2loop2pt}).
\\
\begin{eqnarray}
\text{Moduli space}\ \mathcal{M}_{2,2,0} &:=& 6g-6+2n=10\ \text{real i.e. 5 complex dimensions},\\
\text{Required no. of PCOs} &:=& 2g-2+n=4\ .
\end{eqnarray} 
We call the coordinate on $S_1$ as $z_1$, on $S_2$ as $z_2$, on $S_3$ as $z'_2$ and on $S_4$ as $z'_1$. The coordinate on the disk $D_1$ around the first puncture is $w_1$ such that the location of the puncture in this coordinate is $w_1=0$. Similarly, the coordinate on the disk $D_2$ around the second puncture is $w_2$ with
the location of the puncture at $w_2=0$. We depict these choice of coordinates on the right in figure \ref{fig:2loop2pt}. 

With this choice of local coordinates we write down the transition functions as,
\begin{eqnarray}
z_1\ =\ w_1+\zeta_1 && \text{on $C_1$},\label{eq:cp1}\\
z_1\ =\ z_2 && \text{on $C_3$},\\
z_2\ =\ -\frac{q}{z'_2} && \text{on $C_5$ (with $|q|\geq 1$)},\\
z'_2\ =\ z'_1 && \text{on $C_6$},\\
z'_1\ =\ w_2+\zeta_2 && \text{on $C_2$},\\
z_1\ =\ z_2-\tau_1 && \text{on $C_4$},\\
z'_2\ =\ z'_1-\tau_2 && \text{on $C_7$}\ .\label{eq:cp2}
\end{eqnarray}
Here $\tau_1,\tau_2,\zeta_1,\zeta_2,q$ are 5 complex parameters accounting for the 5 complex moduli that are required. Let us remark here once more that the fibre directions over the base moduli space also contain the locations of the PCOs. The violet points denote the locations of the 4 PCOs at $y_1,y_2,y_3$ and $y_4$. As in the 1 point case we take $y_1\rightarrow p_1$ and $y_2\rightarrow p_2$ at the very end.

\subsection{The 2 point function as a total derivative}
\label{sec:g2p2_2}
The Beltrami differentials in this case are,
\begin{eqnarray}
B_{\tau_1}=-\oint_{C_4}b(z_1)dz_1, && B_{\bar{\tau}_1}=-\oint_{C_4}\bar{b}(\bar{z}_1)d\bar{z}_1\\
B_{\tau_2}=-\oint_{C_7}b(z'_2)dz'_2, && B_{\bar{\tau}_2}=-\oint_{C_7}\bar{b}(\bar{z}'_2)d\bar{z}'_2\\
B_{\zeta_1}=\oint_{C_1}b(w_1)dw_1, && B_{\bar{\zeta}_1}=\oint_{C_1}\bar{b}(\bar{w}_1)d\bar{w}_1\\
B_{\zeta_2}=\oint_{C_2}b(w_2)dw_2, && B_{\bar{\zeta}_2}=\oint_{C_2}\bar{b}(\bar{w}_2)d\bar{w}_2\\
B_{q}=\oint_{C_5}q^{-1}z_2b(z_2)dz_2, && B_{\bar{q}}=\oint_{C_5}\bar{q}^{-1}\bar{z}_2\bar{b}(\bar{z}_2)d\bar{z}_2
\end{eqnarray}
We can now write down the 2 point function on genus 2 Riemann surface with two on-shell external -1 picture vertex which will be our starting point.
\begin{equation}
\begin{aligned}
A_2^{(2)}=& (-2\pi i)^{-5}g_s^4\int d\tau_1\wedge d\bar{\tau}_1\wedge d\tau_2\wedge d\bar{\tau}_2\wedge d\zeta_1\wedge d\bar{\zeta}_1\wedge d\zeta_2\wedge d\bar{\zeta}_2\wedge dq\wedge d\bar{q} \\
&\hspace{1cm}\bigg\langle B_{\tau_1}B_{\bar{\tau}_1}B_{\tau_2}B_{\bar{\tau}_2}B_{\zeta_1}B_{\bar{\zeta}_1}B_{\zeta_2}B_{\bar{\zeta}_2}B_{q}B_{\bar{q}}\ \xi(z_0)\prod_{\alpha=1}^{4}\chi(y_{\alpha})f_1\circ V_1(0)f_2\circ V_2(0)\bigg\rangle\ .
\end{aligned}
\end{equation}
Now we will put one PCO on top of the -1 picture vertex $V_1$ to convert it to a 0 picture vertex and keep the other one as it is. The reason for this is we want to pick up the residue for all the spurious poles i.e. even the ones which may depend on $\zeta_2$. Following the previous section we then write the 0 picture vertex as,
$$V_1^{(0)}(\zeta_1)=\oint_{\zeta_1} dx \mathcal{J}(x)V_1^{(\frac{1}{2})}(\zeta_1)$$ 
where the 10d spacetime supercurrent $\mathcal{J}(x)$ is already defined in the previous section.

Again we pull the $x$ contour out and close it at some other point away from $\zeta_1$ and $\zeta_2$ picking up the contributions from the 8 spurious poles $r'_l$ in the process. So we have,
\begin{equation}
\begin{aligned}
A_2^{(2)}&= (-2\pi i)^{-5}g_s^4\int d\tau_1\wedge d\bar{\tau}_1\wedge d\tau_2\wedge d\bar{\tau}_2\wedge d\zeta_1\wedge d\bar{\zeta}_1\wedge d\zeta_2\wedge d\bar{\zeta}_2\wedge dq\wedge d\bar{q} \\
&\sum_{r'_l}\oint_{r'_l}\frac{dx}{2\pi i}\bigg\langle B_{\tau_1}B_{\bar{\tau}_1}B_{\tau_2}B_{\bar{\tau}_2}B_{\zeta_1}B_{\bar{\zeta}_1}B_{\zeta_2}B_{\bar{\zeta}_2}B_{q}B_{\bar{q}}\ \xi(z_0)\prod_{\alpha=1}^{3}\chi(y_{\alpha})\mathcal{J}(x)V_1^{(\frac{1}{2})}(\zeta_1,\bar{\zeta}_1)V_2(\zeta_2,\bar{\zeta}_2)\bigg\rangle\ .
\end{aligned}
\end{equation}
We can see clearly that the location of the spurious poles depend on $y_1,y_2,y_3,\zeta_1$ and $\zeta_2$. By moving the PCO at $y_1$ to a suitable location $\tilde{y}_1$ we can ensure that none of the new poles coincide with $r'_l$ any longer. As a result the contour integrals around $r'_l$ with the PCO at $\tilde{y}_1$ simply vanishes and thus we can rewrite the integrand as,
\begin{equation}
\sum_{r'_l}\oint_{r'_l}\frac{dx}{2\pi i}\bigg\langle B_{\tau_1}\dots B_{\bar{q}}\ \xi(z_0)\lbrace Q_B,(\xi(y_1)-\xi(\tilde{y}_1))\rbrace\prod_{\alpha=2}^{3}\chi(y_{\alpha})\mathcal{J}(x)V_1^{(\frac{1}{2})}(\zeta_1,\bar{\zeta}_1)V_2(\zeta_2,\bar{\zeta}_2)\bigg\rangle\ .
\end{equation}
Since the external states are on-shell and $\xi(y_1)-\xi(\tilde{y}_1)$ is an operator in the small Hilbert space, we can use \eqref{eq:main_id} once again to write this integrand as a total derivative and thus represent it as a boundary term, viz.   
$$\lim_{a\rightarrow 0}\int_0^{2\pi}d\theta\ (\bar{q}F'(q,\bar{q}))|_{r=a}\ .$$
Taking $z_0=\tilde{y}_1$, the quantity $F'(q,\bar{q})$ is given by,
\begin{equation}
\begin{aligned}
&F'(q,\bar{q})=\\
&(-2\pi i)^{-5}g_s^4\int d\tau_1\wedge d\bar{\tau}_1\wedge d\tau_2\wedge d\bar{\tau}_2\wedge d\zeta_1\wedge d\bar{\zeta}_1\wedge d\zeta_2\wedge d\bar{\zeta}_2\\
&\sum_{r'_l}\oint_{r'_l}\frac{dx}{2\pi i}\bigg\langle B_{\tau_1}B_{\bar{\tau}_1}B_{\tau_2}B_{\bar{\tau}_2}B_{\zeta_1}B_{\bar{\zeta}_1}B_{\zeta_2}B_{\bar{\zeta}_2}B_{\bar{q}}\ \xi(\tilde{y}_1)\xi(y_1)\prod_{\alpha=2}^{3}\chi(y_{\alpha})\mathcal{J}(x)V_1^{(\frac{1}{2})}(\zeta_1,\bar{\zeta}_1)V_2(\zeta_2,\bar{\zeta}_2)\bigg\rangle\ .
\end{aligned}
\end{equation}

\subsection{Contribution due to the $\xi,\eta,\phi$ and the $b,c$ ghost}
\label{sec:g2p2_3}
As described in the case of 1 point function most of the contributions vanish due to $\phi$ charge conservation. The only non vanishing piece is,
\begin{equation}
\begin{aligned}
&F'(q,\bar{q})=\\
&-\frac{g_s^4}{4}(-2\pi i)^{-5}\bigg[\lim_{y_2\rightarrow y'_2}\int d\tau_1\wedge d\bar{\tau}_1\wedge d\tau_2\wedge d\bar{\tau}_2\wedge d\zeta_1\wedge d\bar{\zeta}_1\wedge d\zeta_2\wedge d\bar{\zeta}_2 \left(2\frac{\partial}{\partial y_2}+\frac{\partial}{\partial y'_2}\right)G'(y_2,y'_2)\\
&\ \ +\lim_{y_3\rightarrow y'_3}\int d\tau_1\wedge d\bar{\tau}_1\wedge d\tau_2\wedge d\bar{\tau}_2\wedge d\zeta_1\wedge d\bar{\zeta}_1\wedge d\zeta_2\wedge d\bar{\zeta}_2 \left(2\frac{\partial}{\partial y_3}+\frac{\partial}{\partial y'_3}\right)G'(y_3,y'_3)\bigg]
\end{aligned}
\end{equation}
where, 
\begin{equation}
\begin{aligned}
G'(y_2,y'_2)=&\sum_{r'_l}\oint_{r'_l}\frac{dx}{2\pi i}\bigg\langle B_{\tau_1}B_{\bar{\tau}_1}B_{\tau_2}B_{\bar{\tau}_2}B_{\zeta_1}B_{\bar{\zeta}_1}B_{\zeta_2}B_{\bar{\zeta}_2}B_{\bar{q}}\\
&\hspace{1cm} \xi(\tilde{y}_1)\xi(y_1)\eta(y_2)e^{2\phi(y'_2)}b(y'_2)e^{\phi(y_3)}T_F(y_3)\mathcal{J}(x)V_1^{(\frac{1}{2})}(\zeta_1,\bar{\zeta}_1)V_2(\zeta_2,\bar{\zeta}_2)\bigg\rangle\ 
\end{aligned}
\label{eq:Gy2}
\end{equation}
and
\begin{equation}
\begin{aligned}
G'(y_3,y'_3)=&\sum_{r'_l}\oint_{r'_l}\frac{dx}{2\pi i}\bigg\langle B_{\tau_1}B_{\bar{\tau}_1}B_{\tau_2}B_{\bar{\tau}_2}B_{\zeta_1}B_{\bar{\zeta}_1}B_{\zeta_2}B_{\bar{\zeta}_2}B_{\bar{q}}\\
&\hspace{1cm} \xi(\tilde{y}_1)\xi(y_1)e^{\phi(y_2)}T_F(y_2)\eta(y_3)e^{2\phi(y'_3)}b(y'_3)\mathcal{J}(x)V_1^{(\frac{1}{2})}(\zeta_1,\bar{\zeta}_1)V_2(\zeta_2,\bar{\zeta}_2)\bigg\rangle\ .
\end{aligned}
\label{eq:Gy3}
\end{equation}
We will focus on the details while computing $G'(y_2,y'_2)$ and write down the result for $G'(y_3,y'_3)$ directly since the steps followed are similar.

The superghost part of the correlation function \eqref{eq:Gy2} is given by,
\begin{equation}
\begin{aligned}
&\left\langle \xi(\tilde{y}_1)\xi(y_1)\eta(y_2)e^{2\phi(y'_2)}e^{\phi(y_3)}e^{-\frac{\phi(x)}{2}}e^{\frac{\phi(\zeta_1)}{2}}e^{-\phi(\zeta_2)}\right\rangle =\\
&\frac{(Z_1)^{\frac{1}{2}}\ \vartheta[\delta]\left(\vec{\tilde{y}}_1+\vec{y}_1-2\vec{y}_2+2\vec{y}^{\ \prime}_2-\frac{\vec{x}}{2}+\frac{\vec{\zeta}_1}{2}+\vec{y}_3-\vec{\zeta}_2-2\vec{\Delta}\right)}{\vartheta[\delta]\left(\vec{y}_1-\vec{y}_2+2\vec{y}^{\ \prime}_2-\frac{\vec{x}}{2}+\frac{\vec{\zeta}_1}{2}+\vec{y}_3-\vec{\zeta}_2-2\vec{\Delta}\right)\vartheta[\delta]\left(\vec{\tilde{y}}_1-\vec{y}_2+2\vec{y}^{\ \prime}_2-\frac{\vec{x}}{2}+\frac{\vec{\zeta}_1}{2}+\vec{y}_3-\vec{\zeta}_2-2\vec{\Delta}\right)}\\
&\ \ \ \frac{E(\tilde{y}_1,y_1)E(x,y'_2)(E(x,\zeta_1))^{\frac{1}{4}}(E(x,y_3))^{\frac{1}{2}}(E(y'_2,\zeta_2))^2E(y_3,\zeta_2)}{E(y'_2,\zeta_1)E(\tilde{y}_1,y_2)E(y_1,y_2)(E(y'_2,y_3))^2(E(x,\zeta_2))^{\frac{1}{2}}}\frac{\sigma(x)(\sigma(\zeta_2))^2}{\sigma(\zeta_1)(\sigma(y'_2))^4(\sigma(y_3))^2}\ .
\end{aligned}
\label{eq:Gy2_sg}
\end{equation}
Similarly, from \eqref{eq:Gy3} we get,
\begin{equation}
\begin{aligned}
&\left\langle \xi(\tilde{y}_1)\xi(y_1)\eta(y_3)e^{\phi(y_2)}e^{2\phi(y'_3)}e^{-\frac{\phi(x)}{2}}e^{\frac{\phi(\zeta_1)}{2}}e^{-\phi(\zeta_2)}\right\rangle =\\
&\frac{(Z_1)^{\frac{1}{2}}\ \vartheta[\delta]\left(\vec{\tilde{y}}_1+\vec{y}_1-2\vec{y}_3+2\vec{y}^{\ \prime}_3-\frac{\vec{x}}{2}+\frac{\vec{\zeta}_1}{2}+\vec{y}_2-\vec{\zeta}_2-2\vec{\Delta}\right)}{\vartheta[\delta]\left(\vec{y}_1-\vec{y}_3+2\vec{y}^{\ \prime}_3-\frac{\vec{x}}{2}+\frac{\vec{\zeta}_1}{2}+\vec{y}_2-\vec{\zeta}_2-2\vec{\Delta}\right)\vartheta[\delta]\left(\vec{\tilde{y}}_1-\vec{y}_3+2\vec{y}^{\ \prime}_3-\frac{\vec{x}}{2}+\frac{\vec{\zeta}_1}{2}+\vec{y}_2-\vec{\zeta}_2-2\vec{\Delta}\right)}\\
&\ \ \ \frac{E(\tilde{y}_1,y_1)E(x,y'_3)(E(x,\zeta_1))^{\frac{1}{4}}(E(x,y_2))^{\frac{1}{2}}(E(y'_3,\zeta_2))^2E(y_2,\zeta_2)}{E(y'_3,\zeta_1)E(\tilde{y}_1,y_3)E(y_1,y_3)(E(y'_3,y_2))^2(E(x,\zeta_2))^{\frac{1}{2}}}\frac{\sigma(x)(\sigma(\zeta_2))^2}{\sigma(\zeta_1)(\sigma(y'_3))^4(\sigma(y_2))^2}\ .
\end{aligned}
\label{eq:Gy3_sg}
\end{equation}

\subsubsection{Identifying the 1PI subspace}
Let us now first put $y_1,y_3$ on $T_1$ while $\tilde{y}_1,y_2, y'_2$ are put on $T_2$. Of course $\zeta_1$ and $\zeta_2$ are integrated over so we have to consider different configuration in which they lie on different tori. 

But here we keep in mind that all the regions of integration where $\zeta_1$ and $\zeta_2$ lie on different tori are part of the 1PR subspace i.e. these contributions come from the diagrams of type shown in figure \ref{fig:1PR} and thus we drop these contributions in computing the renormalised mass. So the regions of integration relevant for our calculation are either both $\zeta_1$ and $\zeta_2$ lying on $T_1$ or both lying on $T_2$. We will work with both cases separately.
\subsubsection*{Both $\zeta_1$ and $\zeta_2$ lying on $T_1$}
For $G'(y_2,y'_2)$ we get,
\begin{enumerate}
\item{When $x$ lies on $T_1$: For $q\rightarrow 0$,
\begin{equation}
\begin{aligned}
&\vartheta[\delta]\left(\vec{y}_1-\vec{y}_2+2\vec{y}^{\ \prime}_2-\frac{\vec{x}}{2}+\frac{\vec{\zeta}_1}{2}+\vec{y}_3-\vec{\zeta}_2-2\vec{\Delta}\right)\\
&\rightarrow
\vartheta\begin{bmatrix}
a_1\\
b_1
\end{bmatrix}\left(y_1-\frac{x}{2}+\frac{\zeta_1}{2}+y_3-\zeta_2-p_1|\tau_1\right)
\vartheta\begin{bmatrix}
a_2\\
b_2
\end{bmatrix}\left(2y'_2-y_2-p_2|\tau_2\right)\ .
\end{aligned}
\end{equation}
For a given spin structure $(a_1,b_1)$ the above function vanishes for,
$$
\frac{x}{2}=y_1+\frac{\zeta_1}{2}+y_3-\zeta_2-p_1+\hat{a}_1\tau_1+\hat{b}_1
$$
with four different spin structures $(a_2,b_2)$. The residues at these four poles contain the factor,
\begin{equation}
\begin{aligned}
&\vartheta[\delta]\left(\vec{\tilde{y}}_1+\vec{y}_1-2\vec{y}_2+2\vec{y}^{\ \prime}_2-\frac{\vec{x}}{2}+\frac{\vec{\zeta}_1}{2}+\vec{y}_3-\vec{\zeta}_2-2\vec{\Delta}\right)\\
&\rightarrow
\vartheta\begin{bmatrix}
a_1\\
b_1
\end{bmatrix}\left(y_1-\frac{x}{2}+\frac{\zeta_1}{2}+y_3-\zeta_2-p_1|\tau_1\right)
\vartheta\begin{bmatrix}
a_2\\
b_2
\end{bmatrix}\left(\tilde{y}_1+2y'_2-2y_2-p_2|\tau_2\right)\ .
\end{aligned}
\end{equation}
which vanish exactly at the locations of the poles written above. As a result the contribution from the residues at these four poles vanish.}
\item{When $x$ lies on $T_2$: For $q\rightarrow 0$ we find,
\begin{equation}
\begin{aligned}
&\vartheta[\delta]\left(\vec{y}_1-\vec{y}_2+2\vec{y}^{\ \prime}_2-\frac{\vec{x}}{2}+\frac{\vec{\zeta}_1}{2}+\vec{y}_3-\vec{\zeta}_2-2\vec{\Delta}\right)\\
&\rightarrow
\vartheta\begin{bmatrix}
a_1\\
b_1
\end{bmatrix}\left(y_1+\frac{\zeta_1}{2}+y_3-\zeta_2-\frac{3p_1}{2}|\tau_1\right)
\vartheta\begin{bmatrix}
a_2\\
b_2
\end{bmatrix}\left(2y'_2-y_2-\frac{x}{2}-\frac{p_2}{2}|\tau_2\right)\ .
\end{aligned}
\end{equation}
This function vanishes for a given spin structure $(a_2,b_2)$ at,
$$\frac{x}{2}=2y'_2-y_2-\frac{p_2}{2}+\hat{a}_2\tau_2+\hat{b}_2\ .$$
For 4 different spin structures $(a_1,b_1)$ this accounts for the remaining 4 poles that lie on $T_2$. The superghost contribution to the residues at these poles upto an overall phase are given by,}
\end{enumerate}  
\begin{equation}
\begin{aligned}
&q^{3/8}\eta_d(\tau_1)\eta_d(\tau_2) \frac{(\vartheta'_1\left(0|\tau_1\right))^{-\frac{5}{4}}(\vartheta'_1\left(0|\tau_2\right))^{-\frac{9}{4}}}{\vartheta_{a_1b_1}\left(\frac{\zeta_1}{2}+y_3-\zeta_2-\frac{p_1}{2}|\tau_1\right)}\\
&\quad\times\frac{\vartheta_1\left(x-y'_2|\tau_2\right)(\vartheta_1\left(\zeta_1-p_1|\tau_1\right))^{\frac{1}{4}}(\vartheta_1\left(y_3-p_1|\tau_1\right))^{\frac{1}{2}}(\vartheta_1\left(p_2-y'_2|\tau_2\right))^3\vartheta_1(y_3-\zeta_2|\tau_1)}{(\vartheta_1\left(p_2-x|\tau_2\right))^{\frac{3}{4}}(\vartheta_1\left(\zeta_2-p_1|\tau_1\right))^{\frac{1}{2}}\vartheta_1\left(p_2-y_2|\tau_2\right)}\ .
\end{aligned}
\label{eq:superghost_2pt}
\end{equation}

Now consider the contribution due to $b,c$ ghost correlation function which is exactly the as in the 1 point function i.e. ,
\begin{equation}
\langle B_{\tau_1}B_{\bar{\tau}_1}B_{\tau_2}B_{\bar{\tau}_2}B_{\bar{q}}\ b(y'_2)\rangle=\frac{1}{\bar{q}^{2}q}|\eta_d(\tau_1)|^4|\eta_d(\tau_2)|^4\left(\frac{\vartheta'_1(0|\tau_2)}{\vartheta_1(y'_2-p_2|\tau_2)}\right)^2\ .
\label{eq:bc_2pt}
\end{equation}
Putting \eqref{eq:superghost_2pt} and \eqref{eq:bc_2pt} together, we get the full ghost contribution to the residues to be,
\begin{equation}
\begin{aligned}
&\frac{1}{q^{5/8}\bar{q}^2}(\eta_d(\tau_1)\eta_d(\tau_2))^3(\bar{\eta}_d(\bar{\tau}_1)\bar{\eta}_d(\bar{\tau}_2))^2 \frac{(\vartheta'_1\left(0|\tau_1\right))^{-\frac{5}{4}}(\vartheta'_1\left(0|\tau_2\right))^{-\frac{1}{4}}}{\vartheta_{a_1b_1}\left(\frac{\zeta_1}{2}+y_3-\zeta_2-\frac{p_1}{2}|\tau_1\right)}\\
&\quad\times\frac{\vartheta_1\left(3y'_2-2y_2-p_2|\tau_2\right)(\vartheta_1\left(\zeta_1-p_1|\tau_1\right))^{\frac{1}{4}}(\vartheta_1\left(y_3-p_1|\tau_1\right))^{\frac{1}{2}}\vartheta_1\left(y'_2-p_2|\tau_2\right)\vartheta_1(y_3-\zeta_2|\tau_1)}{(\vartheta_1\left(2p_2+2y_2-4y'_2|\tau_2\right))^{\frac{3}{4}}(\vartheta_1\left(\zeta_2-p_1|\tau_1\right))^{\frac{1}{2}}\vartheta_1\left(y_2-p_2|\tau_2\right)}\ .
\end{aligned}
\label{eq:superghost_2pt1}
\end{equation}

Following similar arguments and steps we can get the non vanishing contributions for $G'(y_3,y'_3)$ from \eqref{eq:Gy3_sg}. The $b,c$ ghost contribution in this case is,
\begin{equation}
\langle B_{\tau_1}B_{\bar{\tau}_1}B_{\tau_2}B_{\bar{\tau}_2}B_{\bar{q}}\ b(y'_3)\rangle=\frac{1}{\bar{q}^{2}q}|\eta_d(\tau_1)|^4|\eta_d(\tau_2)|^4\left(\frac{\vartheta'_1(0|\tau_1)}{\vartheta_1(y'_3-p_1|\tau_1)}\right)^2\ .
\end{equation} 
It is easy to check that the expression \eqref{eq:Gy3_sg} will have contributions from poles on both tori $T_1$ and $T_2$. The residues at the 4 poles on $T_1$ ($\frac{x}{2}=y_1-y_3+2y'_3+\frac{\zeta_1}{2}-\zeta_2-p_1+\hat{a}_1\tau_1+\hat{b}_1$) due to both superghost and $b,c$ ghost comes out to be,
\begin{equation}
\begin{aligned}
&\frac{1}{q^{3/4}\bar{q}^2}(\eta_d(\tau_1)\eta_d(\tau_2))^3(\bar{\eta}_d(\bar{\tau}_1)\bar{\eta}_d(\bar{\tau}_2))^2 \frac{(\vartheta'_1\left(0|\tau_1\right))^{-\frac{1}{4}}(\vartheta'_1\left(0|\tau_2\right))^{-\frac{3}{2}}\vartheta_1\left(3y'_3-2y_3+2y_1+\zeta_1-2\zeta_2-2p_1|\tau_1\right)}{\vartheta_{a_2b_2}\left(y_2-p_2|\tau_2\right)(\vartheta_1\left(4y'_3-2y_3+2y_1-3\zeta_2+\zeta_1-2p_1|\tau_1\right))^{\frac{1}{2}}}\\
&\times\frac{(\vartheta_1\left(4y'_3-2y_3+2y_1-2\zeta_2-2p_1|\tau_1\right))^{\frac{1}{4}}(\vartheta_1\left(p_2-y_2|\tau_2\right))^{\frac{3}{2}}\vartheta_1\left(\zeta_1-p_1|\tau_1\right)(\vartheta_1\left(y'_3-\zeta_2|\tau_1\right))^2}{\vartheta_1\left(y'_3-\zeta_1|\tau_1\right)\vartheta_1\left(y_1-y_3|\tau_1\right)(\vartheta_1\left(4y'_3-2y_3+2y_1-2\zeta_2+\zeta_1-3p_1|\tau_1\right))^{\frac{1}{2}}\vartheta_1\left(\zeta_2-p_1|\tau_1\right)}\ .
\end{aligned}
\label{eq:superghost_2pt2}
\end{equation}\\
The residues for the other 4 poles i.e. those on $T_2$ contains a factor of $\vartheta_1(y_1-p_1|\tau_1)$ in the numerator and hence this will vanish when we take $y_1\rightarrow p_1$ at end of our computation.
\subsubsection*{Both $\zeta_1$ and $\zeta_2$ lying on $T_2$}
It should be clear by now that beginning our analysis from \eqref{eq:Gy2_sg} and \eqref{eq:Gy3_sg} we take different configurations of the location of the vertices and $x$ to figure out the relevant contributions to $G(y_2,y'_2)$ and $G(y_3,y'_3)$ respectively. The procedure remains exactly same as in the previous case so we directly write down the final result for the full ghost contributions.

For $G(y_2,y'_2)$ we get,
\begin{equation}
\begin{aligned}
&\frac{1}{q^{3/4}\bar{q}^2}(\eta_d(\tau_1)\eta_d(\tau_2))^3(\bar{\eta}_d(\bar{\tau}_1)\bar{\eta}_d(\bar{\tau}_2))^2 \frac{(\vartheta'_1\left(0|\tau_1\right))^{-\frac{3}{2}}(\vartheta'_1\left(0|\tau_2\right))^{-\frac{1}{4}}\vartheta_1\left(3y'_2-2y_2+\zeta_1-2\zeta_2|\tau_2\right)}{\vartheta_{a_1b_1}\left(y_3-p_1|\tau_1\right)(\vartheta_1\left(4y'_2-2y_2-3\zeta_2+\zeta_1-2p_1|\tau_1\right))^{\frac{1}{2}}}\\
&\times\frac{(\vartheta_1\left(4y'_2-2y_2-2\zeta_2|\tau_1\right))^{\frac{1}{4}}(\vartheta_1\left(y_3-p_1|\tau_1\right))^{\frac{3}{2}}\vartheta_1\left(\zeta_1-p_2|\tau_2\right)(\vartheta_1\left(y'_2-\zeta_2|\tau_2\right))^2}{\vartheta_1\left(y'_2-\zeta_1|\tau_2\right)\vartheta_1\left(y_2-p_2|\tau_2\right)(\vartheta_1\left(p_2-4y'_2+2y_2+2\zeta_2-\zeta_1|\tau_2\right))^{\frac{1}{2}}\vartheta_1\left(\zeta_2-p_2|\tau_2\right)}\ .
\end{aligned}
\label{eq:superghost_2pt3}
\end{equation}\\
For $G(y_3,y'_3)$ we get (dropping the piece proportional to $\vartheta_1(y_1-p_1|\tau_1)$),
\begin{equation}
\begin{aligned}
&\frac{1}{q^{5/8}\bar{q}^2}(\eta_d(\tau_1)\eta_d(\tau_2))^3(\bar{\eta}_d(\bar{\tau}_1)\bar{\eta}_d(\bar{\tau}_2))^2 \frac{(\vartheta'_1\left(0|\tau_1\right))^{-\frac{1}{4}}(\vartheta'_1\left(0|\tau_2\right))^{-\frac{5}{4}}}{\vartheta_{a_2b_2}\left(\frac{\zeta_1}{2}+y_2-\zeta_2-\frac{p_2}{2}|\tau_2\right)}\\
&\quad\times\frac{\vartheta_1\left(3y'_3-2y_3+2y_1-3p_1|\tau_1\right)(\vartheta_1\left(p_2-\zeta_1|\tau_2\right))^{\frac{1}{4}}(\vartheta_1\left(p_2-y_2|\tau_2\right))^{\frac{1}{2}}\vartheta_1\left(y'_3-p_1|\tau_1\right)\vartheta_1(\zeta_2-y_2|\tau_2)}{(\vartheta_1\left(4y'_3-2y_3+2y_1-4p_1|\tau_1\right))^{\frac{3}{4}}(\vartheta_1\left(p_2-\zeta_2|\tau_2\right))^{\frac{1}{2}}\vartheta_1\left(y_1-y_3|\tau_1\right)}\ .
\end{aligned}
\label{eq:superghost_2pt4}
\end{equation}

With the ghost contribution fully determined we can now turn our attention to the matter sector. We first consider the massless states to show that the renormalised mass vanishes as expected from the non renormalisation theorems and then focus on the massive states whose renormalised mass at this loop order is the new result that we set out to determine.

\subsection{The massless states}
\label{sec:massless}
The vertex operator in different picture for the massless states are already provided in the previous section so we proceed with those expressions to write the correlation function relevant for the current scenario.
\subsubsection*{Both $\zeta_1$ and $\zeta_2$ lying on $T_1$}
The matter sector for $G'(y_2,y'_2)$ is,
\begin{equation}
\begin{aligned}
&u^{\alpha}(k_1)v^{\alpha'}(k_2)(\gamma_{\mu})_{\beta\alpha'}\epsilon_a\tilde{\epsilon}_{\nu b}\bigg\langle \psi^{\rho}(y_3)\partial X_{\rho}(y_3)S_{\alpha}(x)S^{\beta}(\zeta_1)\left(\partial X^{\mu}-\frac{1}{4}i(k_1.\psi)\psi^{\mu}\right)(\zeta_1)\\
&\hspace*{2cm}\bar{J}^a(\bar{\zeta}_1)\psi^{\nu}(\zeta_2)\bar{J}^b(\bar{\zeta}_2)e^{ik_1.X(\zeta_1,\bar{\zeta}_1)}e^{ik_2.X(\zeta_2,\bar{\zeta}_2)}\bigg\rangle\sim (2\pi)^{10}\delta^{10}(k_1+k_2)(\bar{u}\gamma_{\mu}v)\langle\dots\rangle.
\end{aligned}
\end{equation}
We will focus on the fermionic part first. To get the fermionic contribution for,
\begin{equation}
\begin{aligned}
&\bigg\langle \psi^{\rho}(y_3)\psi^{\nu}(\zeta_2)S_{\alpha}(x)S^{\beta}(\zeta_1)\partial X_{\rho}(y_3)\partial X^{\mu}(\zeta_1)\bar{J}^a(\bar{\zeta}_1)\bar{J}^b(\bar{\zeta}_2)e^{ik_1.X(\zeta_1,\bar{\zeta}_1)}e^{-ik_1.X(\zeta_2,\bar{\zeta}_2)}\bigg\rangle \nonumber
\end{aligned}
\end{equation}
we need,
\begin{equation}
\begin{aligned}
\langle \psi^{\rho}(y_3)\psi^{\nu}(\zeta_2)\dots\rangle=&\pm\frac{\eta^{\rho\nu}}{2}\langle(\Psi_j\pm\bar{\Psi}_j)(y_3)(\Psi_j\pm\bar{\Psi}_j)(\zeta_2)\dots\rangle\\
=&\pm\frac{\eta^{\rho\nu}}{2}\langle(\Psi_j(y_3)\bar{\Psi}_j(\zeta_2)+\bar{\Psi}_j(y_3)\Psi_j(\zeta_2))\dots\rangle\ .
\end{aligned}
\label{eq:fermion2pt}
\end{equation}
One can check that the contribution from the other two terms simply vanish due to fermionic charge conservation. Also note that the $+$ sign is for $\rho=\sigma=1,\dots,5$, while $-$ sign is for $\rho=\sigma=6,\dots,10$.
As in the 1 point case we now set, 
$$S_{\alpha}S^{\beta}\equiv \delta_{\alpha}^{\ \beta}S_{1-}(x)\dots S_{5-}(x)S_1^{\ +}(\zeta_1)\dots S_5^{\ +}(\zeta_1)\ $$
for the first term in \eqref{eq:fermion2pt} and
$$S_{\alpha}S^{\beta}\equiv \delta_{\alpha}^{\ \beta}S_{1+}(x)\dots S_{5+}(x)S_1^{\ -}(\zeta_1)\dots S_5^{\ -}(\zeta_1)\ $$ 
for the second term in \eqref{eq:fermion2pt}.
We then apply the factorisation theorem \eqref{eq:factorization} in each case and pick up the piece relevant to get the $1/\bar{q}$ divergent piece for $F'(q,\bar{q})$. For the current case, the relevant operators must have conformal dimension $(1,\frac{5}{8})$ i.e.,
$$\varphi(p_1)\equiv S_{1-}(p_1)S_{2-}(p_1)S_{3-}(p_1)S_{4-}(p_1)S_{5-}(p_1)\bar{J}^b(p_1)\ ,$$
$$\Rightarrow\varphi^{\dagger}(p_2)\equiv S_1^{\ +}(p_2)S_2^{\ +}(p_2)S_3^{\ +}(p_2)S_4^{\ +}(p_2)S_5^{\ +}(p_2)\bar{J}^b(p_2)\ ,$$
for the first and
$$\varphi(p_1)\equiv S_{1+}(p_1)S_{2+}(p_1)S_{3+}(p_1)S_{4+}(p_1)S_{5+}(p_1)\bar{J}^b(p_1)\ ,$$
$$\Rightarrow\varphi^{\dagger}(p_2)\equiv S_1^{\ -}(p_2)S_2^{\ -}(p_2)S_3^{\ -}(p_2)S_4^{\ -}(p_2)S_5^{\ -}(p_2)\bar{J}^b(p_2)\ ,$$
for the second term respectively. So, finally we get,
\begin{equation}
\begin{aligned}
&\sum_c\pm\frac{\eta^{\nu\rho}}{2}q^{5/8}\bar{q}\left\langle \partial X_{\rho}(y_3)\partial X^{\mu}(\zeta_1)\bar{J}^a(\bar{\zeta}_1)\bar{J}^b(\bar{\zeta}_2)\bar{J}^c(p_1)e^{ik_1.X(\zeta_1,\bar{\zeta}_1)}e^{-ik_1.X(\zeta_2,\bar{\zeta}_2)}\right\rangle_{T_1}\langle \bar{J}^c(p_2)\rangle_{T_2}\\
&\bigg[\langle \Psi_j(y_3)\bar{\Psi}_j(\zeta_2)S_1^{\ +}(\zeta_1)S_2^{\ +}(\zeta_1)S_3^{\ +}(\zeta_1)S_4^{\ +}(\zeta_1)S_5^{\ +}(\zeta_1)S_{1-}(p_1)S_{2-}(p_1)S_{3-}(p_1)S_{4-}(p_1)S_{5-}(p_1)\rangle_{a_1b_1}\\
&\langle S_{1-}(x)S_{2-}(x)S_{3-}(x)S_{4-}(x)S_{5-}(x)S_1^{\ +}(p_2)S_2^{\ +}(p_2)S_3^{\ +}(p_2)S_4^{\ +}(p_2)S_5^{\ +}(p_2)\rangle_{a_2b_2}\\
&+\langle \bar{\Psi}_j(y_3)\Psi_j(\zeta_2)S_1^{\ -}(\zeta_1)S_2^{\ -}(\zeta_1)S_3^{\ -}(\zeta_1)S_4^{\ -}(\zeta_1)S_5^{\ -}(\zeta_1)S_{1+}(p_1)S_{2+}(p_1)S_{3+}(p_1)S_{4+}(p_1)S_{5+}(p_1)\rangle_{a_1b_1}\\
&\langle S_{1+}(x)S_{2+}(x)S_{3+}(x)S_{4+}(x)S_{5+}(x)S_1^{\ -}(p_2)S_2^{\ -}(p_2)S_3^{\ -}(p_2)S_4^{\ -}(p_2)S_5^{\ -}(p_2)\rangle_{a_2b_2}\bigg]\ .
\end{aligned}
\label{eq:type1}
\end{equation} 
Summing over the spin structures for the fermionic correlator on $T_1$ after putting it together with the ghost contribution \eqref{eq:superghost_2pt1} we get\footnote{The factor $(-1)^{4a_1b_1}$ occurs due to the reason that $\vartheta_{a_1b_1}(-z|\tau)=(-1)^{4a_1b_1}\vartheta_{a_1b_1}(z|\tau)$ which takes care of the fact that the theta function is odd for $(a_1,b_1)=(1/2,1/2)$ and even otherwise.},
\begin{equation}
\begin{aligned}
&\frac{1}{\bar{q}}\sum_{a_1,b_1}\varepsilon(a_1,b_1)\bigg[\frac{\big(\vartheta_{a_1b_1}\left(\frac{\zeta_1}{2}-\frac{p_1}{2}|\tau_1\right)\big)^4\vartheta_{a_1b_1}\left(\frac{\zeta_1}{2}+y_3-\zeta_2-\frac{p_1}{2}|\tau_1\right)}{\vartheta_{a_1b_1}\left(\frac{\zeta_1}{2}+y_3-\zeta_2-\frac{p_1}{2}|\tau_1\right)}(\dots)\\
&+(-1)^{4a_1b_1}\frac{\big(\vartheta_{a_1b_1}\left(\frac{p_1}{2}-\frac{\zeta_1}{2}|\tau_1\right)\big)^4\vartheta_{a_1b_1}\left(\frac{\zeta_1}{2}+y_3-\zeta_2-\frac{p_1}{2}|\tau_1\right)}{\vartheta_{a_1b_1}\left(\frac{\zeta_1}{2}+y_3-\zeta_2-\frac{p_1}{2}|\tau_1\right)}(\dots)\bigg]=0\ ,
\end{aligned}
\end{equation}
by using \eqref{eq:theta_id}. The ($\dots$) denotes the rest of the pices and they are independent of $(a_1,b_1)$

Next we have,
\begin{equation}
\begin{aligned}
&-\frac{i}{4}k_{1\sigma}\bigg\langle \psi^{\rho}(y_3)\psi^{\sigma}(\zeta_1)\psi^{\mu}(\zeta_1)\psi^{\nu}(\zeta_2)S_{\alpha}(x)S^{\beta}(\zeta_1)\partial X_{\rho}(y_3)\bar{J}^a(\bar{\zeta}_1)\bar{J}^b(\bar{\zeta}_2)e^{ik_1.X(\zeta_1,\bar{\zeta}_1)}e^{-ik_1.X(\zeta_2,\bar{\zeta}_2)}\bigg\rangle\ .\nonumber
\end{aligned}
\end{equation}
To compute the fermionic part in this case let us first determine the possible tensor structures,
\begin{equation}
\begin{aligned}
\langle \psi^{\rho}(y_3)\psi^{\sigma}(\zeta_1)\psi^{\mu}(\zeta_1)\psi^{\nu}(\zeta_2)\dots\rangle=&A_1(\zeta_1,\zeta_2,y_3)\eta^{\rho\sigma}\eta^{\mu\nu}+A_2(\zeta_1,\zeta_2,y_3)\eta^{\rho\mu}\eta^{\sigma\nu}\\
&+A_3(\zeta_1,\zeta_2,y_3)\eta^{\rho\nu}\eta^{\mu\sigma}+A_4(\zeta_1,\zeta_2,y_3)\eta^{\rho\nu}\Sigma^{\mu\sigma}
\end{aligned}
\label{eq:f_tensor1}
\end{equation}
while from the bosonic part we have,
\begin{equation}
\bigg\langle \partial X_{\rho}(y_3)e^{ik_1.X(\zeta_1,\bar{\zeta}_1)}e^{-ik_1.X(\zeta_2,\bar{\zeta}_2)}\dots\bigg\rangle\sim ik_{1\rho}\left(\frac{1}{2E(y_3,\zeta_2)}-\frac{1}{2E(y_3,\zeta_1)}\right)\ .
\end{equation}
Now it is easy to check that all the terms in \eqref{eq:f_tensor1} simply vanishes due to the on-shell conditions,
\begin{equation}
k_1^2=0\ ,\quad \slashed{k}_1v=0\ ,\quad\text{and}\quad k_1^{\nu}\tilde{\epsilon}_{\nu b}=-k_2^{\nu}\tilde{\epsilon}_{\nu b}=0\ .
\label{eq:eom_2pt}
\end{equation}

In a similar fashion one can check that for the matter sector of $G'(y_3,y'_3)$, 
\begin{equation}
\begin{aligned}
&u^{\alpha}(k_1)v^{\alpha'}(k_2)(\gamma_{\mu})_{\beta\alpha'}\epsilon_a\tilde{\epsilon}_{\nu b}\bigg\langle \psi^{\rho}(y_2)\partial X_{\rho}(y_2)S_{\alpha}(x)S^{\beta}(\zeta_1)\left(\partial X^{\mu}-\frac{1}{4}i(k_1.\psi)\psi^{\mu}\right)(\zeta_1)\\
&\hspace*{2cm}\bar{J}^a(\bar{\zeta}_1)\psi^{\nu}(\zeta_2)\bar{J}^b(\bar{\zeta}_2)e^{ik_1.X(\zeta_1,\bar{\zeta}_1)}e^{ik_2.X(\zeta_2,\bar{\zeta}_2)}\bigg\rangle\sim (2\pi)^{10}\delta^{10}(k_1+k_2)(\bar{u}\gamma_{\mu}v)\langle\dots\rangle\ ,
\end{aligned}
\end{equation}
the term with four fermions vanishes due to the on-shell conditions \eqref{eq:eom_2pt}. For the term with two fermions it is a bit more subtle. The operators which are relevant in the factorisation for this case must have conformal weight $(1,\frac{3}{4})$. The fact that we need the fermion charge to be conserved to get a non zero result leads to the fact that among all operators of weight $(1,\frac{3}{4})$ the only ones we should focus on are,
$$\phi(p_1)=S_{j}^{\ +}(p_1)S_{j -}(p_1)\Psi_{j}(p_1)\bar{J}^b(p_1)\ ,\quad S_{j}^{\ +}(p_1)S_{j -}(p_1)\bar{\Psi}_{j}(p_1)\bar{J}^b(p_1)\ ,$$
$$\Rightarrow\phi^{\dagger}(p_2)=S_{j-}(p_2)S_{j}^{\ +}(p_2)\bar{\Psi}_{j}(p_2)\bar{J}^b(p_2)\ ,\quad S_{j-}(p_2)S_{j}^{\ +}(p_2)\bar{\Psi}_{j}(p_2)\bar{J}^b(p_2)\ .$$
As a result we end up with,
\begin{equation}
\begin{aligned}
&\sum_c\pm\frac{\eta^{\nu\rho}}{2}q^{3/4}\bar{q}\left\langle \partial X^{\mu}(\zeta_1)\bar{J}^a(\bar{\zeta}_1)\bar{J}^b(\bar{\zeta}_2)\bar{J}^c(p_1)e^{ik_1.X(\zeta_1,\bar{\zeta}_1)}e^{-ik_1.X(\zeta_2,\bar{\zeta}_2)}\right\rangle_{T_1}\langle \partial X_{\rho}(y_2)\bar{J}^c(p_2)\rangle_{T_2}\\
&\bigg[\langle \Psi_j(y_2)\bar{\Psi}_{j}(p_2)S_{j-}(p_2)S_{j}^{\ +}(p_2)\rangle_{a_2b_2}\langle \bar{\Psi}_j(\zeta_2)\Psi_{j}(p_1)S_{j}^{\ +}(p_1)S_{j -}(p_1)\\
&\hspace*{3cm}S_1^{\ +}(\zeta_1)S_2^{\ +}(\zeta_1)S_3^{\ +}(\zeta_1)S_4^{\ +}(\zeta_1)S_5^{\ +}(\zeta_1)S_{1-}(x)S_{2-}(x)S_{3-}(x)S_{4-}(x)S_{5-}(x)\rangle_{a_1b_1}\\
&+\langle \bar{\Psi}_j(y_2)\Psi_{j}(p_2)S_{j-}(p_2)S_{j}^{\ +}(p_2)\rangle_{a_2b_2}\langle \Psi_j(\zeta_2)\bar{\Psi}_{j}(p_1)S_{j}^{\ +}(p_1)S_{j -}(p_1)\\
&\hspace*{3cm}S_1^{\ +}(\zeta_1)S_2^{\ +}(\zeta_1)S_3^{\ +}(\zeta_1)S_4^{\ +}(\zeta_1)S_5^{\ +}(\zeta_1)S_{1-}(x)S_{2-}(x)S_{3-}(x)S_{4-}(x)S_{5-}(x)\rangle_{a_1b_1}\bigg]\ .
\end{aligned}
\label{eq:type1}
\end{equation} 
After putting this together with \eqref{eq:superghost_2pt2}, taking contribution from all the fermions and summing over the spin structures on $T_2$ we get,
\begin{equation}
\begin{aligned}
&\frac{1}{\bar{q}}\sum_{a_2,b_2}\varepsilon(a_2,b_2)\bigg[\frac{\big(\vartheta_{a_2b_2}\left(0|\tau_2\right)\big)^4\vartheta_{a_2b_2}\left(y_2-p_2|\tau_2\right)}{\vartheta_{a_2b_2}\left(y_2-p_2|\tau_2\right)}(\dots)\\
&+(-1)^{4a_2b_2}\frac{\big(\vartheta_{a_2b_2}\left(0|\tau_2\right)\big)^4\vartheta_{a_2b_2}\left(y_2-p_2|\tau_2\right)}{\vartheta_{a_2b_2}\left(y_2-p_2|\tau_2\right)}(\dots)\bigg]=0\ ,
\end{aligned}
\end{equation}
by using \eqref{eq:theta_id}. The ($\dots$) denotes the rest of the pices which are independent of $(a_2,b_2)$.\subsubsection*{Both $\zeta_1$ and $\zeta_2$ lying on $T_2$}
The analysis in this case is exactly the same. The only thing to note is that the expressions for $G'(y_2,y'_2)$ and $G'(y_3,y'_3)$ get interchanged and the ghost parts relevant in this case are respectively \eqref{eq:superghost_2pt3} and \eqref{eq:superghost_2pt4}. So $G'(y_2,y'_2)$ and $G'(y_3,y'_3)$ vanish in this case as well. Of course at the end we have to take $y_1\rightarrow p_1$ as well as $y_2\rightarrow p_2$. As a result the terms proportional to $\vartheta_1(y_1-p_1|\tau_1)$ and $\vartheta_1(y_1-p_1|\tau_1)$ goes to zero like we mentioned earlier. So finally we have,
$$G'(y_2,y'_2)=G'(y_3,y'_3)=0\ ,\quad\Rightarrow F'(q,\bar{q})=0$$ 
inside the regions of integration where both $\zeta_1$ and $\zeta_2$ lie on $T_1$ and where both lie on $T_2$ i.e. the 1PI subspace. So the renormalised mass of the massless string states vanish at two loop as expected.

\subsection{The massive states}
\label{sec:massive}
Let us first write down the vertex operators for a state at first massive level in $-1$ and $1/2$ picture since these are the ones we will need for our computation. Since, this is a massive state we can work in the rest frame of the string i.e. $k_{\mu}\equiv (k_0,0,0,\dots)$ so that in the $\alpha'=1$ unit we have,
$$\text{(mass)}^2=-k^2=(k_0)^2=4$$  
the vertex operators in this frame are given by,
\begin{eqnarray}
\text{-1 picture}:&& V_j(z,\bar{z})\ =\ \epsilon_{ab}c\bar{c}e^{-\phi(z)}\psi^1\psi^2\psi^3(z)\bar{J}^a\bar{J}^b(\bar{z})e^{ik_{j0}X^0(z,\bar{z})}\\
\text{1/2 picture}:&& V_j^{(\frac{1}{2})}(z,\bar{z})\ =\ \frac{i}{\sqrt{2}}\epsilon_{ab}v^{\alpha}e^{ik_{j0}X^0(z,\bar{z})}\bar{J}^a\bar{J}^b(\bar{z})\bigg\lbrace e^{\frac{\phi(z)}{2}}\bigg(\psi^1\psi^2\partial X^3 \nonumber\\
&&\quad\ +\ \psi^2\psi^3\partial X^1+\psi^3\psi^1\partial X^2-ik_{j0}\psi^1\psi^2\psi^3\psi^0\bigg)(\gamma_{123})_{\beta\alpha}S^{\beta}(z)+\dots\bigg\rbrace.
\end{eqnarray}
The $\dots$ piece will not be needed for the computation since we only need the single pole contribution when acted on by $\mathcal{J}(x)$, but we know that this piece has $\phi$ charge 3/2.

\subsubsection*{Both $\zeta_1$ and $\zeta_2$ lying on $T_1$}
With the above vertex operators we now note first that $G'(y_2,y'_2)$ has two types of terms viz.,
\begin{equation}
\begin{aligned}
&-u^{\alpha}(k_{10})v^{\alpha'}(k_{20})(\gamma_{123})_{\beta\alpha'}\epsilon_{ab}\tilde{\epsilon}_{cd}\bigg\langle \psi^{\rho}(y_3)\psi^{1}(\zeta_1)\psi^{2}(\zeta_1)\psi^{1}(\zeta_2)\psi^{2}(\zeta_2)\psi^{3}(\zeta_2)S_{\alpha}(x)S^{\beta}(\zeta_1)\\
&\hspace*{3cm}\partial X_{\rho}(y_3)\partial X^{3}(\zeta_1)\bar{J}^a(\bar{\zeta}_1)\bar{J}^b(\bar{\zeta}_1)\bar{J}^c(\bar{\zeta}_2)\bar{J}^d(\bar{\zeta}_2)e^{ik_{10}X^0(\zeta_1,\bar{\zeta}_1)}e^{ik_{20}X^0(\zeta_2,\bar{\zeta}_2)}\bigg\rangle\\
&\hspace*{8cm}\sim 2\pi\delta(k_{10}+k_{20})(\bar{u}\gamma_{123}v)\epsilon_{ab}\tilde{\epsilon}_{cd}\langle\dots\rangle\ ,
\end{aligned}
\label{eq:massive_t1}
\end{equation}
and,
\begin{equation}
\begin{aligned}
&ik_{10}u^{\alpha}(k_{10})v^{\alpha'}(k_{20})(\gamma_{123})_{\beta\alpha'}\epsilon_{ab}\tilde{\epsilon}_{cd}\bigg\langle \psi^{\rho}(y_3)\psi^{1}(\zeta_1)\psi^{2}(\zeta_1)\psi^{3}(\zeta_1)\psi^{0}(\zeta_1)\psi^{1}(\zeta_2)\psi^{2}(\zeta_2)\psi^{3}(\zeta_2)\\
&\hspace*{3cm}S_{\alpha}(x)S^{\beta}(\zeta_1)\partial X_{\rho}(y_3)\bar{J}^a(\bar{\zeta}_1)\bar{J}^b(\bar{\zeta}_1)\bar{J}^c(\bar{\zeta}_2)\bar{J}^d(\bar{\zeta}_2)e^{ik_{10}X^0(\zeta_1,\bar{\zeta}_1)}e^{ik_{20}X^0(\zeta_2,\bar{\zeta}_2)}\bigg\rangle\\
&\hspace*{8cm}\sim 2\pi\delta(k_{10}+k_{20})(\bar{u}\gamma_{123}v)\epsilon_{ab}\tilde{\epsilon}_{cd}\langle\dots\rangle\ .
\end{aligned}
\label{eq:massive_t2}
\end{equation}
Of course there are three terms of the first type obtained by cyclically permuting 1,2 and 3. Owing to the factorisation theorem \eqref{eq:factorization} we see that just like in the massless case the relevant operators for these correlation functions must have conformal weight $(1,\frac{5}{8})$ and hence should be of the form,
$$\varphi(p_1)\equiv S_{1\pm}(p_1)S_{2\pm}(p_1)S_{3\pm}(p_1)S_{4\pm}(p_1)S_{5\pm}(p_1)\bar{J}^b(p_1)\ ,$$
$$\Rightarrow\varphi^{\dagger}(p_2)\equiv S_1^{\ \pm}(p_2)S_2^{\ \pm}(p_2)S_3^{\ \pm}(p_2)S_4^{\ \pm}(p_2)S_5^{\ \pm}(p_2)\bar{J}^b(p_2)\ ,$$

Let us first analyse the type \eqref{eq:massive_t1}. From the bosonic part we can see that the index $\rho$ can be either 3 or 0. The readers can convince themselves that when $\rho=0$ (using $\psi^0=i\psi^{10}$) the fermionic correlation becomes such that if we try to conserve the fermion charge on $T_1$ the fermion charge of $T_2$ gets violated and vice versa. As a result the contribution for $\rho=0$ must vanish. So we are left with $\rho=3$ and this implies that,
\begin{equation}
\begin{aligned}
\langle\psi^{3}(y_3)\psi^{1}(\zeta_1)\psi^{2}(\zeta_1)\psi^{1}(\zeta_2)\psi^{2}(\zeta_2)\psi^{3}(\zeta_2)..\rangle &=-\langle\psi^{1}(\zeta_1)\psi^{1}(\zeta_2)\psi^{2}(\zeta_1)\psi^{2}(\zeta_2)\psi^{3}(y_3)\psi^{3}(\zeta_2)..\rangle\\
&=-\frac{1}{8}\bigg(\langle\Psi_{1}(\zeta_1)\bar{\Psi}_{1}(\zeta_2)\Psi_{2}(\zeta_1)\bar{\Psi}_{2}(\zeta_2)\Psi_{3}(y_3)\bar{\Psi}_{3}(\zeta_2)..\rangle\\
& +\langle\bar{\Psi}_{1}(\zeta_1)\Psi_{1}(\zeta_2)\Psi_{2}(\zeta_1)\bar{\Psi}_{2}(\zeta_2)\Psi_{3}(y_3)\bar{\Psi}_{3}(\zeta_2)..\rangle\\
& +\langle\Psi_{1}(\zeta_1)\bar{\Psi}_{1}(\zeta_2)\bar{\Psi}_{2}(\zeta_1)\Psi_{2}(\zeta_2)\Psi_{3}(y_3)\bar{\Psi}_{3}(\zeta_2)..\rangle\\
&+\dots+\langle\bar{\Psi}_{1}(\zeta_1)\Psi_{1}(\zeta_2)\bar{\Psi}_{2}(\zeta_1)\Psi_{2}(\zeta_2)\bar{\Psi}_{3}(y_3)\Psi_{3}(\zeta_2)..\rangle\bigg)
\end{aligned}
\end{equation}
where the $\dots$ denote 4 more terms. Now, we have $\langle..\Psi_3(y_3)\bar{\Psi}_3(\zeta_2)..\rangle$ for four of these terms for which we set,
$$S_{\alpha}(x)S^{\beta}(\zeta_1)\equiv \delta_{\alpha}^{\ \beta}S_{1-}(x)S_{2-}(x)S_{3-}(x)S_{4-}(x)S_{5-}(x)S_{1}^{+}(\zeta_1)S_{2}^{+}(\zeta_1)S_{3}^{+}(\zeta_1)S_{4}^{+}(\zeta_1)S_{5}^{+}(\zeta_1),$$
while for the other four terms with $\langle..\bar{\Psi}_3(y_3)\Psi_3(\zeta_2)..\rangle$, we set,
$$S_{\alpha}(x)S^{\beta}(\zeta_1)\equiv \delta_{\alpha}^{\ \beta}S_{1-}(x)S_{2-}(x)S_{3+}(x)S_{4-}(x)S_{5-}(x)S_{1}^{+}(\zeta_1)S_{2}^{+}(\zeta_1)S_{3}^{-}(\zeta_1)S_{4}^{+}(\zeta_1)S_{5}^{+}(\zeta_1),$$
The operators of conformal weight $(1,\frac{5}{8})$ relevant in these cases are respectively,
$$\varphi(p_1)\equiv S_{1-}(p_1)S_{2-}(p_1)S_{3-}(p_1)S_{4-}(p_1)S_{5-}(p_1)\bar{J}^e(p_1)\ ,$$
$$\Rightarrow\varphi^{\dagger}(p_2)\equiv S_1^{+}(p_2)S_2^{+}(p_2)S_3^{+}(p_2)S_4^{+}(p_2)S_5^{+}(p_2)\bar{J}^e(p_2)$$
and
$$\varphi(p_1)\equiv S_{1-}(p_1)S_{2-}(p_1)S_{3+}(p_1)S_{4-}(p_1)S_{5-}(p_1)\bar{J}^e(p_1)\ ,$$
$$\Rightarrow\varphi^{\dagger}(p_2)\equiv S_1^{+}(p_2)S_2^{+}(p_2)S_3^{-}(p_2)S_4^{+}(p_2)S_5^{+}(p_2)\bar{J}^e(p_2)\ .$$
For both these cases we focus only on the fermionic part of the correlation function. For the first case there are 4 terms of which we illustrate the first one only. For the rest we will write the final result which can be easily checked.
\begin{equation}
\begin{aligned}
&q^{5/8}\bar{q}\langle\Psi_{1}(\zeta_1)\bar{\Psi}_{1}(\zeta_2)\Psi_{2}(\zeta_1)\bar{\Psi}_{2}(\zeta_2)\Psi_{3}(y_3)\bar{\Psi}_{3}(\zeta_2)S_{1}^{+}(\zeta_1)S_{2}^{+}(\zeta_1)S_{3}^{+}(\zeta_1)S_{4}^{+}(\zeta_1)S_{5}^{+}(\zeta_1)\\
&S_{1-}(p_1)S_{2-}(p_1)S_{3-}(p_1)S_{4-}(p_1)S_{5-}(p_1)\rangle_{a_1b_1}\\
&\langle S_{1-}(x)S_{2-}(x)S_{3-}(x)S_{4-}(x)S_{5-}(x)S_1^{+}(p_2)S_2^{+}(p_2)S_3^{+}(p_2)S_4^{+}(p_2)S_5^{+}(p_2)\rangle_{a_2b_2}
\end{aligned}
\end{equation}
The $\bar{J}^e$ fields are omitted from the expression since they contribute to the bosonic part. Putting the ghost contribution \eqref{eq:superghost_2pt1} with this, the correlation function on $T_1$ yields,
\begin{equation}
\frac{1}{\bar{q}}\sum_{a_1,b_1}\varepsilon(a_1,b_1)\vartheta_{a_1b_1}\left(\frac{3\zeta_1}{2}-\zeta_2-\frac{p_1}{2}|\tau_1\right)^2\vartheta_{a_1b_1}\left(\frac{\zeta_1}{2}-\frac{p_1}{2}|\tau_1\right)^2(\dots)=0
\end{equation}
using \eqref{eq:theta_id}. It is easy to see that if we have $\bar{\Psi}_1(\zeta_1)\Psi_1(\zeta_2)$ instead (keeping everything else unchanged) we end up with,
\begin{equation}
\frac{1}{\bar{q}}\sum_{a_1,b_1}\varepsilon(a_1,b_1)\vartheta_{a_1b_1}\left(-\frac{\zeta_1}{2}+\zeta_2-\frac{p_1}{2}|\tau_1\right)\vartheta_{a_1b_1}\left(\frac{3\zeta_1}{2}-\zeta_2-\frac{p_1}{2}|\tau_1\right)\vartheta_{a_1b_1}\left(\frac{\zeta_1}{2}-\frac{p_1}{2}|\tau_1\right)^2(\dots)=0
\end{equation}
again by the use of \eqref{eq:theta_id}. The readers can now convince themselves that all the other terms also vanish and in exactly the same way the 4 terms of the other case ($\bar{\Psi}_3(y_3)\Psi_3(\zeta_2)$) vanishes as well by repeated use of \eqref{eq:theta_id}. Of course in that case there will be an overall factor of $(-1)^{4a_1b_1}$ but it doesn't matter since the result vanishes anyway.  

Let us now focus on the type \eqref{eq:massive_t2}. We set,
$$S_{\alpha}(x)S^{\beta}(\zeta_1)\equiv \delta_{\alpha}^{\beta}S_{1-}(x)S_{2-}(x)S_{3-}(x)S_{4-}(x)S_{5-}(x)S_{1}^{+}(\zeta_1)S_{2}^{+}(\zeta_1)S_{3}^{+}(\zeta_1)S_{4}^{+}(\zeta_1)S_{5}^{+}(\zeta_1)$$
and thus the relevant operator of weight $(1,\frac{5}{8})$ is,
$$\varphi(p_1)\equiv S_{1-}(p_1)S_{2-}(p_1)S_{3-}(p_1)S_{4-}(p_1)S_{5-}(p_1)\bar{J}^e(p_1)\ ,$$
$$\Rightarrow\varphi^{\dagger}(p_2)\equiv S_1^{+}(p_2)S_2^{+}(p_2)S_3^{+}(p_2)S_4^{+}(p_2)S_5^{+}(p_2)\bar{J}^e(p_2).$$
With this let us first evaluate the bosonic part of the correlation function.
\begin{equation}
\begin{aligned}
&\sum_e\bigg\langle\partial X_{\rho}(y_3)\bar{J}^a(\bar{\zeta}_1)\bar{J}^b(\bar{\zeta}_1)\bar{J}^c(\bar{\zeta}_2)\bar{J}^d(\bar{\zeta}_2)\bar{J}^e(p_1)e^{ik_{10}X^0(\zeta_1,\bar{\zeta}_1)}e^{-ik_{10}X^0(\zeta_2,\bar{\zeta}_2)}\bigg\rangle_{T_1}\bigg\langle\bar{J}^e(p_2)\bigg\rangle_{T_2}\\
&=-\frac{ik_{10}\delta_{\rho}^{\ 0}C(\tau_1)}{2}\left(\frac{\vartheta'_1(0|\tau_1)}{\vartheta_1(y_3-\zeta_1|\tau_1)}-\frac{\vartheta'_1(0|\tau_1)}{\vartheta_1(y_3-\zeta_2|\tau_1)}\right)\bigg[e^{\left\lbrace-2\pi(k_{10})^2\frac{(\text{Im}(\zeta_1-\zeta_2))^2}{\text{Im}\tau_1}\right\rbrace}\bigg|\frac{\vartheta_1(\zeta_1-\zeta_2|\tau_1)}{\vartheta'_1(0|\tau_1)}\bigg|^{2(k_{10})^2}\bigg]\\
&\quad\bigg[\frac{(\delta^{bc}\langle\bar{J}^d(p_1)\rangle+\delta^{bd}\langle\bar{J}^c(p_1)\rangle)\langle\bar{J}^a(p_2)\rangle+(\delta^{ac}\langle\bar{J}^d(p_1)\rangle+\delta^{ad}\langle\bar{J}^c(p_1)\rangle)\langle\bar{J}^b(p_2)\rangle}{(\bar{\vartheta}_1(\bar{\zeta}_1-\bar{\zeta}_2|\bar{\tau}_1))^2(\bar{\vartheta}_1(\bar{\zeta}_1-p_1|\bar{\tau}_1))^2}\\
&\quad\ \ +\frac{(\delta^{ad}\langle\bar{J}^b(p_1)\rangle+\delta^{bd}\langle\bar{J}^a(p_1)\rangle)\langle\bar{J}^c(p_2)\rangle+(\delta^{ac}\langle\bar{J}^b(p_1)\rangle+\delta^{bc}\langle\bar{J}^a(p_1)\rangle)\langle\bar{J}^d(p_2)\rangle}{(\bar{\vartheta}_1(\bar{\zeta}_1-\bar{\zeta}_2|\bar{\tau}_1))^2(\bar{\vartheta}_1(\bar{\zeta}_2-p_1|\bar{\tau}_1))^2}\bigg]\\
&\quad\bigg(\sum_e\frac{(\delta^{ac}\delta^{bd}+\delta^{ad}\delta^{bc})\langle\bar{J}^e(p_1)\rangle\langle\bar{J}^e(p_2)\rangle}{(\bar{\vartheta}_1(\bar{\zeta}_1-\bar{\zeta}_2|\bar{\tau}_1))^4}\bigg)\ .
\end{aligned}
\label{eq:massive_bosonic}
\end{equation}
Due to the $\delta_{\rho}^{\ 0}$ in the bosonic piece it should suffice to take the $\rho=0$ contribution for fermionic correlator and thus we get,
\begin{equation}
\begin{aligned}
&\langle\psi^{0}(y_3)\psi^{1}(\zeta_1)\psi^{2}(\zeta_1)\psi^{3}(\zeta_1)\psi^{0}(\zeta_1)\psi^{1}(\zeta_2)\psi^{2}(\zeta_2)\psi^{3}(\zeta_2)\dots\rangle \\
&=\langle\psi^{0}(y_3)\psi^{0}(\zeta_1)\psi^{1}(\zeta_1)\psi^{1}(\zeta_2)\psi^{2}(\zeta_1)\psi^{2}(\zeta_2)\psi^{3}(\zeta_1)\psi^{3}(\zeta_2)\dots\rangle \\
&=-\langle\psi^{10}(y_3)\psi^{10}(\zeta_1)\psi^{1}(\zeta_1)\psi^{1}(\zeta_2)\psi^{2}(\zeta_1)\psi^{2}(\zeta_2)\psi^{3}(\zeta_1)\psi^{3}(\zeta_2)\dots\rangle \\
&=\frac{1}{16}\bigg(\langle\Psi_{5}(y_3)\bar{\Psi}_{5}(\zeta_1)\Psi_{1}(\zeta_1)\bar{\Psi}_{1}(\zeta_2)\Psi_{2}(\zeta_1)\bar{\Psi}_{2}(\zeta_2)\Psi_{3}(\zeta_1)\bar{\Psi}_{3}(\zeta_2)\dots\rangle\\
&\ \ +\langle\Psi_{5}(y_3)\bar{\Psi}_{5}(\zeta_1)\bar{\Psi}_{1}(\zeta_1)\Psi_{1}(\zeta_2)\Psi_{2}(\zeta_1)\bar{\Psi}_{2}(\zeta_2)\Psi_{3}(\zeta_1)\bar{\Psi}_{3}(\zeta_2)\dots\rangle\\
&\ \ +\dots+\langle\bar{\Psi}_{5}(y_3)\Psi_{5}(\zeta_1)\bar{\Psi}_{1}(\zeta_1)\Psi_{1}(\zeta_2)\bar{\Psi}_{2}(\zeta_1)\Psi_{2}(\zeta_2)\bar{\Psi}_{3}(\zeta_1)\Psi_{3}(\zeta_2)\dots\rangle\bigg)\ .
\end{aligned}
\end{equation}
So we see that we need the contribution from 16 terms of which let us evaluate the correlation function for the first one in detail. The others can be written down in a similar fashion.
\begin{equation}
\begin{aligned}
&\frac{q^{5/8}\bar{q}}{16}\langle\Psi_{5}(y_3)\bar{\Psi}_{5}(\zeta_1)\Psi_{1}(\zeta_1)\bar{\Psi}_{1}(\zeta_2)\Psi_{2}(\zeta_1)\bar{\Psi}_{2}(\zeta_2)\Psi_{3}(\zeta_1)\bar{\Psi}_{3}(\zeta_2)S_{1}^{+}(\zeta_1)S_{2}^{+}(\zeta_1)S_{3}^{+}(\zeta_1)S_{4}^{+}(\zeta_1)S_{5}^{+}(\zeta_1)\\
&S_{1-}(p_1)S_{2-}(p_1)S_{3-}(p_1)S_{4-}(p_1)S_{5-}(p_1)\rangle_{a_1b_1}\\
&\langle S_{1-}(x)S_{2-}(x)S_{3-}(x)S_{4-}(x)S_{5-}(x)S_1^{+}(p_2)S_2^{+}(p_2)S_3^{+}(p_2)S_4^{+}(p_2)S_5^{+}(p_2)\rangle_{a_2b_2}\nonumber
\end{aligned}
\end{equation}
\begin{equation}
\begin{aligned}
=&\frac{q^{5/8}\bar{q}}{16}\frac{\tilde{K}(\tau_1)\tilde{K}(\tau_2)}{(\eta_d(\tau_1)\eta_d(\tau_2))^{5}}\left[\frac{(\vartheta'_1(0|\tau_1))^{\frac{21}{4}}}{(\vartheta_1(\zeta_1-p_1|\tau_1))^{\frac{5}{4}}(\vartheta_1(\zeta_1-\zeta_2|\tau_1))^{\frac{3}{2}}(\vartheta_1(y_3-\zeta_1|\tau_1))^{\frac{1}{2}}}\right]\\
&\varepsilon(a_1,b_1)\vartheta_{a_1b_1}\left(\frac{3\zeta_1}{2}-\zeta_2-\frac{p_1}{2}|\tau_1\right)^3\vartheta_{a_1b_1}\left(\frac{\zeta_1}{2}-\frac{p_1}{2}|\tau_1\right)\vartheta_{a_1b_1}\left(y_3-\frac{\zeta_1}{2}-\frac{p_1}{2}|\tau_1\right)\\
&\times\left(\vartheta_{1}\left(p_2+y_2-2y'_2|\tau_2\right)\right)^5\left(\frac{\vartheta'_1(0|\tau_2)}{\vartheta_1(2p_2+2y_2-4y'_2|\tau_2)}\right)^{\frac{5}{4}}\ .
\end{aligned}
\label{eq:massive_fermionic}
\end{equation}
Here we have used the fact that in this case the poles in $x$ plane for $G'(y_2,y'_2)$ occurs at $\frac{x}{2}=2y'_2-y_2-\frac{p_2}{2}+\hat{a}_2\tau_2+\hat{b_2}$. We have to now put together the parts \eqref{eq:superghost_2pt1}, \eqref{eq:massive_bosonic} and \eqref{eq:massive_fermionic} and then sum over the spin structures $(a_1,b_1)$. doing this we get some non-zero result. But now we have to evaluate,
$$\lim_{y'_2\rightarrow y_2}\left(2\frac{\partial}{\partial y_2}+\frac{\partial}{\partial y'_2}\right)G'(y_2,y'_2),\quad\text{and then take $y_1\rightarrow p_1$ and $y_2\rightarrow p_2$}.$$
From the $y_2, y'_2$ dependent part we get,
\begin{equation}
\begin{aligned}
&\bigg[\frac{\vartheta(p_2+y_2-2y'_2|\tau_2)^5}{\vartheta_1(2p_2+2y_2-4y'_2|\tau_2)^2\vartheta_1(y_2-p_2|\tau_2)}\bigg(-\vartheta'_1(3y'_2-2y_2-p_2|\tau_2)\vartheta_1(y'_2-p_2|\tau_2)\\
&+\vartheta_1(3y'_2-2y_2-p_2|\tau_2)\vartheta'_1(y'_2-p_2|\tau_2)\bigg)\\
&-2\vartheta'_1(y_2-p_2|\tau_2)\frac{\vartheta_1(p_2+y_2-2y'_2|\tau_2)^5\vartheta_1(3y'_2-2y_2-p_2|\tau_2)\vartheta_1(y'_2-p_2|\tau_2)}{\vartheta_1(2p_2+2y_2-4y'_2|\tau_2)^2\vartheta_1(y_2-p_2|\tau_2)^2}\bigg]\ .
\end{aligned}
\end{equation}
In the limit $y'_2\rightarrow y_2$ the first term i.e. the first two lines above vanishes and we are left with,
$$-2\vartheta'_1(y_2-p_2|\tau_2)\frac{\vartheta_1(p_2-y_2|\tau_2)^5}{\vartheta_1(2p_2-2y_2|\tau_2)^2}$$
for which,
$$\lim_{y_2\rightarrow p_2} -2\vartheta'_1(y_2-p_2|\tau_2)\frac{(p_2-y_2)^5(\vartheta'_1(0|\tau_2)+(p_2-y_2)\vartheta''_1(0|\tau_2)/2+\dots)^5}{4(p_2-y_2)^2(\vartheta'_1(0|\tau_2)+(2p_2-2y_2)\vartheta''_1(0|\tau_2)/2+\dots)^2}=0$$

While computing $G'(y_3,y'_3)$, we can simply follow the analysis of the massless case and argue that due to the sum over the spin structures $(a_2,b_2)$ on $T_2$ the contribution in this case vanishes. Since both $\zeta_1$ and $\zeta_2$ are on $T_1$, the correlation function on $T_2$ in this case remains the same as in the massless case.

\subsubsection*{Both $\zeta_1$ and $\zeta_2$ lying on $T_2$}
The analysis for $G'(y_2,y'_2)$ mirrors the one for $G'(y_3,y'_3)$ in the previous case and the same argument shows that this contribution vanishes when we sum over the spin structures $(a_1,b_1)$. As for $G'(y_3,y'_3)$ one can check that we immediately get the matter correlation result by interchanging $y_2\leftrightarrow y_3$, $p_1\leftrightarrow p_2$, $\tau_1\leftrightarrow\tau_2$ and taking $y'_2\rightarrow y'_3$ and the poles in $x$ plane to be at $\frac{x}{2}=y_1+2y'_3-y_3-\frac{3p_1}{2}+\hat{a}_1\tau_1+\hat{b_1}$. This implies that the $\vartheta_1(p_2-y_2|\tau_2)^{\frac{1}{2}}$ piece in the numerator \eqref{eq:superghost_2pt4} does not cancel and hence in the limit $y_2\rightarrow p_2$ this contribution will also vanish. 

Although it seems from the above computation that the mass renormalisation i.e. $\delta m_R$ for the massive states also vanish, but note that we have considered the 10d critical string theory in the flat background which is a free theory hence the above result is not in contradiction. When we consider interacting cases i.e. compactified theories as in \cite{Atick:1987qy} we will get something non-zero.

\subsection{The range of integration for the modular parameters $\tau_1$ and $\tau_2$}
\label{sec:MCG}
\subsubsection{The Mapping Class Group on genus 2 Riemann surface}
\begin{figure}[h!]
\centering
\includegraphics[scale=0.3]{"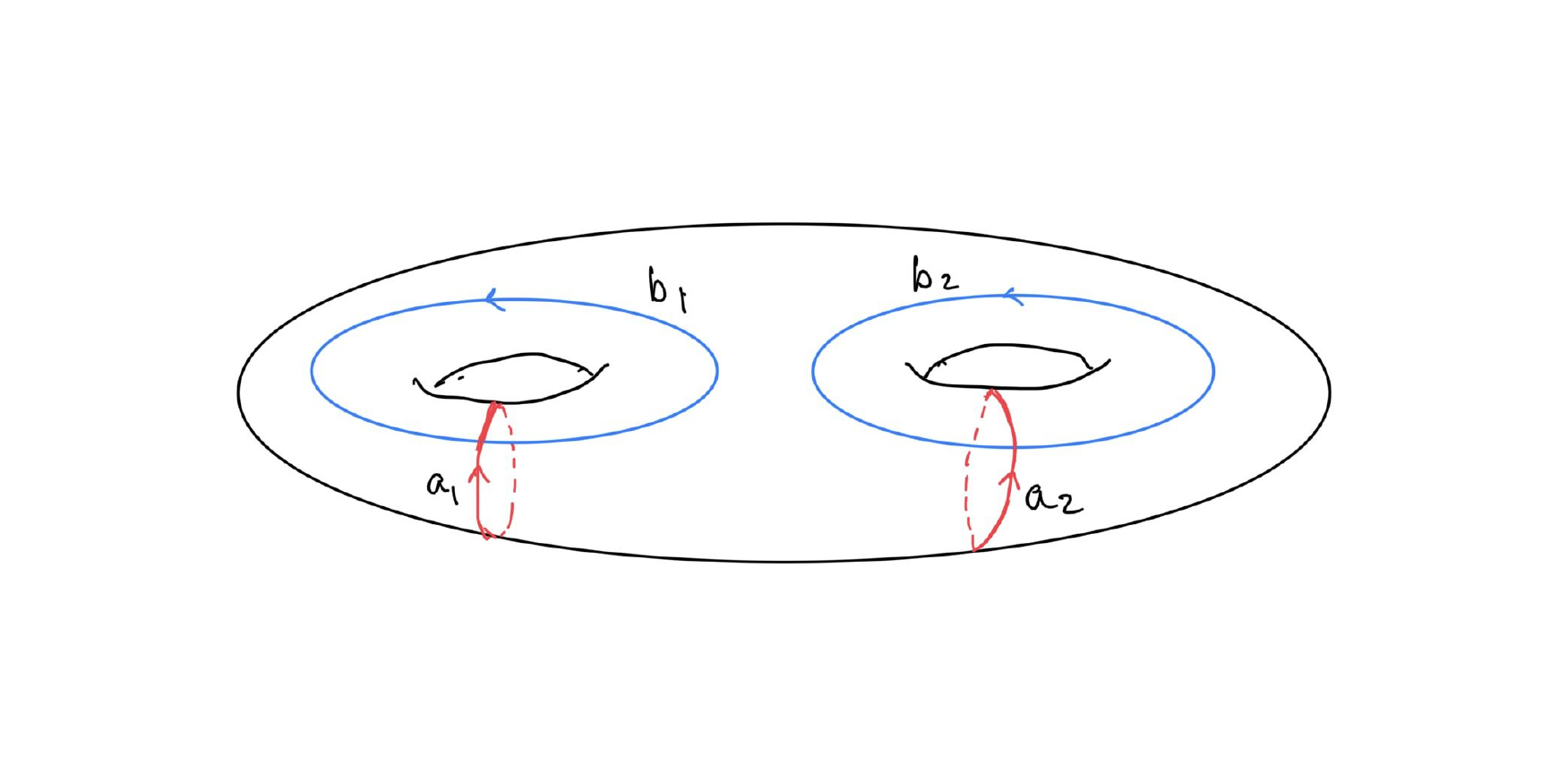"}
\caption{Canonical homology cycles}
\label{fig:cycles}
\end{figure}

In this section we discuss the action of the Mapping Class Group(MCG) on the genus 2 Teichmuller space. We follow the notation and conventions of \cite{Alvarez-Gaume:1986rcs} for this purpose. The MCG for the genus 2 Riemann surface is given by Sp(4,$\mathbb{Z}$). Let us introduce the canonical homology cycles $(a_1,b_1)$ and $(a_2,b_2)$ (figure \ref{fig:cycles}) and the normalised abelian differentials $\omega_1$ and $\omega_2$ such that,
\begin{equation}
\int_{a_i}\omega_j=\delta_{ij}\ ,\quad \int_{b_i}\omega_j=\Omega_{ij}\quad i,j=1,2
\label{eq:abeldiff}
\end{equation}
where $\Omega_{11}=\tau_1$, $\Omega_{22}=\tau_2$ and $\Omega_{12}=\Omega_{21}=q$. As in \cite{Alvarez-Gaume:1986rcs} we consider the following generators of Sp(4,$\mathbb{Z}$),
\begin{equation}
\begin{aligned}
&D_{a_1}=\begin{pmatrix}
1 & 0 & 0 & 0\\
0 & 1 & 0 & 0\\
1 & 0 & 1 & 0\\
0 & 0 & 0 & 1
\end{pmatrix}\ ,\quad 
D_{b_1}=\begin{pmatrix}
1 & 0 & 1 & 0\\
0 & 1 & 0 & 0\\
0 & 0 & 1 & 0\\
0 & 0 & 0 & 1
\end{pmatrix}\ ,\\
&D_{a_2}=\begin{pmatrix}
1 & 0 & 0 & 0\\
0 & 1 & 0 & 0\\
0 & 0 & 1 & 0\\
0 & 1 & 0 & 1
\end{pmatrix}\ ,\quad 
D_{b_1}=\begin{pmatrix}
1 & 0 & 0 & 0\\
0 & 1 & 0 & 1\\
0 & 0 & 1 & 0\\
0 & 0 & 0 & 1
\end{pmatrix}\\
&\quad\quad D_{a_1^{-1}a_2}=\begin{pmatrix}
\ \ 1 & \ \ 0 & 0 & 0\\
\ \ 0 & \ \ 1 & 0 & 0\\
-1 & \ \ 1 & 1 & 0\\
\ \ 1 & -1 & 0 & 1
\end{pmatrix}\ .
\end{aligned}
\label{eq:g2gen}
\end{equation}
The generators have the form of 
$$T=\begin{pmatrix}
D & C\\
B & A
\end{pmatrix}\in \text{Sp(4,$\mathbb{Z}$)},\quad \Rightarrow\ \ A^TD-C^TB=\mathbf{1}_{2\times 2}\ .$$ 
Under these transformations the abelian differentials transform in the following way to keep the normalisation condition \eqref{eq:abeldiff} invariant.
\begin{equation}
\tilde{\omega}_i=\omega_j(C\Omega+D)^{-1}_{ji}\ .
\end{equation} 
As a result the period matrix $\Omega$ transforms as,
\begin{equation}
\Omega\rightarrow \tilde{\Omega}=(A\Omega+B)(C\Omega+D)^{-1},\quad\text{where}\ \ \tilde{\Omega}_{ij}=\int_{b_i}\tilde{\omega}_j
\end{equation}

\subsubsection{A subgroup of the full MCG}
In the course of our calculation we have established that the two loop amplitudes can be written as a total derivative on the moduli space. As a result the full contribution is basically a boundary term, so it suffices to get the range of integration of the modular parameters at the boundary of the moduli space. The boundary of the moduli space of the genus 2 surface is obtained by setting $\Omega_{12}=\Omega_{21}=q=0$. Thus we have at the boundary $$\Omega=\begin{pmatrix}
\tau_1 & 0 \\
0 & \tau_2
\end{pmatrix}\ .$$ 
One can easily check that the 4 generators $D_{a_1},D_{a_2},D_{b_1},D_{b_2}$ keep the point $q=0$ unchanged since,
\begin{eqnarray}
\text{under $D_{a_1}$:}\ \ \Omega\rightarrow\Omega' = \begin{pmatrix}
\tau_1+1 & 0 \\
0 & \tau_2
\end{pmatrix}&,&\text{under $D_{a_2}$:}\ \ \Omega\rightarrow\Omega' = \begin{pmatrix}
\tau_1 & 0 \\
0 & \tau_2+1
\end{pmatrix}\nonumber\\
\text{under $D_{b_1}$:}\ \ \Omega\rightarrow\Omega' = \begin{pmatrix}
\frac{\tau_1}{\tau_1+1} & 0 \\
0 & \tau_2
\end{pmatrix}&,&\text{under $D_{b_2}$:}\ \ \Omega\rightarrow\Omega' = \begin{pmatrix}
\tau_1 & 0 \\
0 & \frac{\tau_2}{\tau_2+1}
\end{pmatrix}\ .
\end{eqnarray}
Hence these 4 generators form a subgroup of Sp($4,\mathbb{Z}$). From the transformations above we can conclude that the parameters $\tau_1$ and $\tau_2$ are those of the two tori $T_1$ and $T_2$ respectively. As in the case of torus we can check that the following boundaries,
\begin{equation}
\left(|\text{Re}(\tau_1)|=\frac{1}{2}\ ,\ |\tau_1|= 1\ ,\ |\text{Re}(\tau_2)|=\frac{1}{2}\ ,\ |\tau_2|= 1\right)
\end{equation} 
are mapped to each other under $D_{a_1},D_{a_2},D_{b_1},D_{b_2}$. This implies that they constitute the boundaries of fundamental region of integration for the boundary($q=0$) of the moduli space of genus 2 Riemann surface. Hence we have range of integration for the parameters $\tau_1$ and $\tau_2$ given by,
\begin{equation}
\mathcal{F}^{(2)}_{q=0}:=\left\lbrace \left(|\text{Re}(\tau_1)|\leq\frac{1}{2},\ |\tau_1|\geq 1\right)\cup\left(|\text{Re}(\tau_2)|\leq\frac{1}{2},\ |\tau_2|\geq 1\right) \right\rbrace\ .
\end{equation}

\section{Discussion}
\label{sec:discuss}
In this work we have provided the relevant integrands and the regions of integration which determines the renormalised mass at two loop order. Of course the expressions for the integrands does depend on the theory under consideration but it differs for the matter sector only. The superghost and the conformal ghost contributions determined in this work remains the same for all theories. It should also be emphasised that the result for massless 1 point and 2 point functions are well known in the literature \cite{Atick:1986es} and we have reproduced those results from the computation method used in \cite{Atick:1987qy}. So they serve as a check of our calculation before we use it to obtain the renormalised mass for the massive states. In this paper we use string theory in 10d flat background which is basically a free theory so we see no renormalised mass for the massive states in this case but we note that the result does not vanish identically. As a result of having a free theory they vanish only at the end when the marked point on the tori $T_1$ and $T_2$ approach two of the PCO locations (one on each torus). For interacting i.e. compactified theories the result approaches a finite non zero answer which depend on the details of the theory. Before ending this article let us mention two points which we were unable to address in more detail. The first is that we work with the same assumption made in \cite{Atick:1987qy} regarding the absence of global obstructions to writing the amplitudes as a total derivative on the moduli space without further inspection. The second is that due to the theta function expression appearing in the spin structure sum for the renormalised mass for the massive states, we were unable to argue that the result is independent of the PCO location $y_3$. This is important since it is part of the spurious data and should not appear for observables with on-shell states.

\section*{Acknowledgements}
I would like to thank Prof. Ashoke Sen for his lectures on ``Superstring Perturbation Theory" in ICTS, Bangalore, March 2022 \cite{Sen:SPT22}, a version of which was also given in LACES, GGI Florence, December 2021 \cite{Sen:SPT21}. Also for clarifying some critical doubts outside the course duration. I would also like to thank Oscar Varela for some useful discussions and comments. This work was supported by the NSF grant PHY-2014163.

\appendix
\section{Theta function Relations}
\label{app:1}
\subsection{Transformation rules}
There are theta functions corresponding to four different spin structures on the torus. They are given by,
\begin{eqnarray}
\vartheta_1(z|\tau)&=&-\vartheta_{\frac{1}{2}\frac{1}{2}}(z|\tau)=-i\sum_{k\in\mathbb{Z}}(-1)^kq^{(k+\frac{1}{2})^2}e^{\pi i(2k+1)z}\ ,\\
\vartheta_2(z|\tau)&=&\vartheta_{\frac{1}{2}0}(z|\tau)=\sum_{k\in\mathbb{Z}}q^{(k+\frac{1}{2})^2}e^{\pi i(2k+1)z}\ ,\\
\vartheta_3(z|\tau)&=&\vartheta_{00}(z|\tau)=\sum_{k\in\mathbb{Z}}q^{k^2}e^{2\pi ikz}\ ,\\
\vartheta_4(z|\tau)&=&\vartheta_{0\frac{1}{2}}(z|\tau)=\sum_{k\in\mathbb{Z}}(-1)^kq^{k^2}e^{2\pi ikz}\ .
\end{eqnarray}
where $q:=e^{i\pi\tau},\ |q|<1$. 
The first one is an odd function of $z$, while the other three are even. The transformation rules under shift by the periods are as follows ($m,n\in\mathbb{Z}$),
\begin{eqnarray}
\vartheta_1(z+m+n\tau|\tau)&=&(-1)^{m+n}e^{-n\pi i(2z+\tau)}\vartheta_{1}(z|\tau)\ ,\\
\vartheta_2(z+m+n\tau|\tau)&=&(-1)^{m}e^{-n\pi i(2z+\tau)}\vartheta_{2}(z|\tau)\ ,\\
\vartheta_3(z+m+n\tau|\tau)&=&e^{-n\pi i(2z+\tau)}\vartheta_{3}(z|\tau)\ ,\\
\vartheta_4(z+m+n\tau|\tau)&=&(-1)^{n}e^{-n\pi i(2z+\tau)}\vartheta_{4}(z|\tau)\ ,
\end{eqnarray}
whereas shift by half periods are,
\begin{eqnarray}
\vartheta_1\left(z+\frac{1}{2}|\tau\right)=\vartheta_2\left(z|\tau\right) &,& \vartheta_1\left(z+\frac{\tau}{2}|\tau\right)=ie^{-\pi i(z+\tau/4)}\vartheta_4\left(z|\tau\right)\ ,\\
\vartheta_2\left(z+\frac{1}{2}|\tau\right)=-\vartheta_1\left(z|\tau\right) &,& \vartheta_2\left(z+\frac{\tau}{2}|\tau\right)=e^{-\pi i(z+\tau/4)}\vartheta_3\left(z|\tau\right)\ ,\\
\vartheta_3\left(z+\frac{1}{2}|\tau\right)=\vartheta_4\left(z|\tau\right) &,& \vartheta_3\left(z+\frac{\tau}{2}|\tau\right)=e^{-\pi i(z+\tau/4)}\vartheta_2\left(z|\tau\right)\ ,\\
\vartheta_4\left(z+\frac{1}{2}|\tau\right)=\vartheta_3\left(z|\tau\right) &,& \vartheta_4\left(z+\frac{\tau}{2}|\tau\right)=ie^{-\pi i(z+\tau/4)}\vartheta_1\left(z|\tau\right)\ .
\end{eqnarray}

\subsection{In the degeneration limit}
For this purpose let us first define the theta function on the genus two surface with spin structure $$\delta=\begin{pmatrix}
a_1 & a_2\\
b_1 & b_2
\end{pmatrix}\equiv
\begin{pmatrix}
\vec{a}\\
\vec{b}
\end{pmatrix}\ ,\quad \text{where $a_i\in\left\lbrace 0,\frac{1}{2}\right\rbrace$ and $b_i\in\left\lbrace 0,\frac{1}{2}\right\rbrace,\ \forall i=1,2$},
$$
as,
\begin{equation}
\vartheta\begin{bmatrix}
\vec{a}\\
\vec{b}
\end{bmatrix}(\vec{z}|\Omega)=\sum_{\vec{n}\in \mathbb{Z}^2}\exp\left(i\pi(\vec{n}+\vec{a})^T.\Omega.(\vec{n}+\vec{a})+2\pi i(\vec{n}+\vec{a})^T.(\vec{z}+\vec{b})\right)\ .
\end{equation}
With this definition one also has the prime form as defined in \eqref{eq:prime_form}.
In the degeneration limit i.e. $q\rightarrow 0$, we have the following results as was given in the appendix of \cite{Atick:1987qy}. For $x_1\in T_1$ and $x_2\in T_2$,
\begin{eqnarray}
E(x_1,x_2)&\rightarrow& q^{-\frac{1}{2}}\left(\frac{\vartheta_1(x_1-p_1|\tau_1)}{\vartheta'_1(0|\tau_1)}\right)\left(\frac{\vartheta_1(p_2-x_2|\tau_2)}{\vartheta'_1(0|\tau_0)}\right)\\
\nonumber\\
E(x_i,x'_i)&\rightarrow& \left(\frac{\vartheta_1(x_i-x'_i|\tau_i)}{\vartheta'_1(0|\tau_i)}\right)\\
\nonumber\\
\left[\delta\right]\equiv\begin{bmatrix}
a_1 & a_2\\
b_1 & b_2
\end{bmatrix}&\rightarrow&
\begin{bmatrix}
a_1 \\
b_1 
\end{bmatrix}+\begin{bmatrix}
a_2\\
b_2
\end{bmatrix}
\end{eqnarray}
\begin{equation}
\begin{aligned}
&\vartheta[\delta]\left(\sum_{i=1}^m\vec{x}_i-\sum_{j=1}^n\vec{y}_j+(m-n)\vec{\Delta}|\Omega\right)\\
&\rightarrow \vartheta
\begin{bmatrix}
a_1 \\
b_1 
\end{bmatrix}\left(\sum_{i=1}^m\vec{x}_i-mp_1-\frac{1}{2}(m-n)(1+\tau_1)|\tau_1\right)\vartheta
\begin{bmatrix}
a_2 \\
b_2 
\end{bmatrix}\left(np_2-\sum_{j=1}^n\vec{y}_j-\frac{1}{2}(m-n)(1+\tau_2)|\tau_2\right)
\end{aligned}
\end{equation}
\begin{eqnarray}
\sigma(x_i)&\rightarrow&\frac{\vartheta'_1(0|\tau_i)}{\vartheta_1(x_i-p_i|\tau_i)}\\
\nonumber\\
Z_1^{\frac{1}{2}}&\rightarrow&\eta_d(\tau_1)\eta_d(\tau_2)\\
\nonumber\\
\omega^1(z|q)&\rightarrow&\left\lbrace \begin{matrix}
1+O(q)\ ,\quad & z\in T_1 \\
O(q)\ ,\quad & z\in T_2
\end{matrix}\right. \\
\omega^2(z|q)&\rightarrow&\left\lbrace \begin{matrix}
O(q)\ ,\quad & z\in T_1 \\
1+O(q)\ ,\quad & z\in T_2
\end{matrix}\right.\ .
\end{eqnarray}
Here $p_1$ and $p_2$ are nodes on $T_1$ and $T_2$ respectively, $\omega^i$ are the abelian differentials and $\eta_d(\tau)$ is the Dedekind eta function. In the above degeneration limit result for the theta function, there seems to be an ambiguity on the right hand side in defining the argument of the theta functions since they can be shifted by integral multiples of 1 and $\tau_1$ (or 1 and $\tau_2$). This reflects the ambiguity in the choice of path from the base point in defining $\vec{x}=\int_{P_0}^P\vec{\omega}$.


\begin{thebibliography}{99}

\bibitem{Nelson:1988ic}
P. C. Nelson, “Covariant Insertion of General Vertex Operators,” Phys. Rev. Lett. 62,
993 (1989).

\bibitem{Vafa:1987es}
C. Vafa, “Operator Formulation on Riemann Surfaces,” Phys. Lett. B 190, 47 (1987).

\bibitem{Vafa:1987ea}   
   C. Vafa, “Conformal Theories and Punctured Surfaces,” Phys. Lett. B 199, 195 (1987).

\bibitem{Cohen:1985sm}
A. G. Cohen, G. W. Moore, P. C. Nelson and J. Polchinski, “An Off-Shell Propagator for String Theory,” Nucl. Phys. B 267, 143 (1986); “Semi Off-shell String Am- plitudes,” Nucl. Phys. B 281, 127 (1987).

\bibitem{Alvarez-Gaume:1987eux}
L. Alvarez Gaume, C. Gomez, G. Moore and C. Vafa, “Strings in the operator formalism”, Nucl. Phys. B 303, 455 (1988).

\bibitem{Alvarez-Gaume:1988lei}
 L. Alvarez Gaume, C. Gomez, P. Nelson, G. Sierra and C. Vafa, “Fermionic strings in the operator formalism”, Nucl. Phys. B 311, 333 (1988).
 
 \bibitem{Polchinski:1988jq}
J. Polchinski, “Factorization of Bosonic String Amplitudes,” Nucl. Phys. B 307, 61 (1988).

\bibitem{Pius:2013sca}
R.~Pius, A.~Rudra and A.~Sen, “Mass Renormalization in String Theory: Special States,” doi:10.1007/JHEP07(2014)058 [arXiv:1311.1257 [hep-th]].

\bibitem{Pius:2014iaa}
R. Pius, A. Rudra and A. Sen, “Mass Renormalization in String Theory: General States,” arXiv:1401.7014 [hep-th].

\bibitem{Witten:1985cc}
E. Witten, “Noncommutative Geometry and String Field Theory,” Nucl. Phys. B 268, 253 (1986).
   
\bibitem{Zwiebach:1992ie}
 B.~Zwiebach, ``Closed string field theory: Quantum action and the B-V master equation", Nucl. Phys. B, doi:10.1016/0550-3213(93)90388-6 [arXiv: hep-th/9206084].
 
\bibitem{Rastelli:2007gg}
L. Rastelli and B. Zwiebach, “The Off-shell Veneziano amplitude in Schnabl gauge,” JHEP 0801, 018 (2008) [arXiv:0708.2591 [hep-th]]. 

\bibitem{Sen:2015uaa}
A.~Sen, ``BV Master Action for Heterotic and Type II String Field Theory",doi: 10.1007/JHEP02(2016)087 JHEP 02, 2016 [arXiv:1508.05387 [hep-th]].

\bibitem{Polchinski:1985zf}
J. Polchinski ``Evaluation of the One Loop String Path Integral", Commun.Math.Phys. 104 (1986) 37.

\bibitem{PhysRevD.5.1945}
Shapiro, J.: Loop graph in the dual-tube model. Phys. Rev. D5, 1945 (1972); Rohm, R.: Spontaneous supersymmetry breaking in supersymmetric string theories. Nucl. Phys. B237,
553 (1984).
	 
\bibitem{Witten:1986qs}
E. Witten, “Interacting Field Theory of Open Superstrings,” Nucl. Phys. B 276, 291 (1986).

\bibitem{Alvarez-Gaume:1986rcs}
Luis Alvarez-Gaume, Gregory W. Moore, Cumrun Vafa ``Theta Functions, Modular Invariance and Strings", Commun.Math.Phys. 106 (1986)

\bibitem{Atick:1986es}
J. J. Atick, A. Sen ``Spin Field Correlators on an Arbitrary Genus Riemann Surface and Nonrenormalization Theorems in String Theories", Phys.Lett.B 186 (1987) 339-346.

\bibitem{Atick:1987qy}
J. J. Atick, A. Sen ``Two Loop Dilaton Tadpole Induced by Fayet-iliopoulos D Terms in Compactified Heterotic String Theories", Nucl.Phys.B 296 (1988) 157-186.

\bibitem{Saroja:1992vw}
R. Saroja and A. Sen, “Picture changing operators in closed fermionic string field theory,” Phys. Lett. B 286, 256 (1992) [hep-th/9202087].

\bibitem{Berkovits:1995ab}
N. Berkovits, “SuperPoincare invariant superstring field theory,” Nucl. Phys. B 450 (1995) 90 [Erratum-ibid. B 459 (1996) 439] [hep-th/9503099].

\bibitem{Berkovits:2001im}
N. Berkovits, “The Ramond sector of open superstring field theory,” JHEP 0111 (2001) 047 [hep-th/0109100].

\bibitem{Okawa:2004ii}
Y. Okawa and B. Zwiebach, “Heterotic string field theory,” JHEP 0407 (2004) 042 [hep-th/0406212].

\bibitem{Berkovits:2004xh}
N. Berkovits, Y. Okawa and B. Zwiebach, “WZW-like action for heterotic string field theory,” JHEP 0411 (2004) 038 [hep-th/0409018].

\bibitem{Kroyter:2012ni}
M. Kroyter, Y. Okawa, M. Schnabl, S. Torii and B. Zwiebach, “Open superstring field theory I: gauge fixing, ghost structure, and propagator,” JHEP 1203 (2012) 030 [arXiv:1201.1761 [hep-th]].

\bibitem{Jurco:2013qra}
B. Jurco and K. Muenster, “Type II Superstring Field Theory: Geometric Approach and Operadic Description,” JHEP 1304 (2013) 126 [arXiv:1303.2323 [hep-th]].

\bibitem{Iimori:2013kha}
Y. Iimori, T. Noumi, Y. Okawa and S. Torii, “From the Berkovits formulation to the Wit- ten formulation in open superstring field theory,” JHEP 1403 (2014) 044 [arXiv:1312.1677 [hep-th]].

\bibitem{Erler:2013xta}
T. Erler, S. Konopka and I. Sachs, “Resolving Witten‘s superstring field theory,” JHEP 1404 (2014) 150 [arXiv:1312.2948 [hep-th]]; “NS-NS Sector of Closed Superstring Field Theory,” arXiv:1403.0940 [hep-th].

\bibitem{Kunitomo:2013mq}
H. Kunitomo, “The Ramond Sector of Heterotic String Field Theory,” PTEP 2014 4, 043B01 [arXiv:1312.7197 [hep-th]].

\bibitem{Kunitomo:2015usa}
H. Kunitomo and Y. Okawa, ``Complete action for open superstring field theory", PTEP 2016 (2016) 2, 023B01 [arXIv: 1508.00366 [hep-th]].

\bibitem{deLacroix:2017lif}
              C.~de Lacroix, H.~Erbin, S.P.~Kashyap, A.~Sen and M.~Verma ``Closed Superstring Field Theory and its Applications'', Int. J. Mod. Phys. A 32, no.28n29, 1730021 (2017) doi:10.1142/S0217751X17300216 [arXiv:1703.06410 [hep-th]].

\bibitem{FarooghMoosavian:2019yke}
S.~F.~Moosavian, A.~Sen, M.~Verma, ``Superstring field theory with open and closed strings", JHEP 01 (2020) 183 [arXiv:1907.10632[hep-th]].

\bibitem{Witten:2012bh}
  E. Witten, “Superstring Perturbation Theory Revisited,” arXiv:1209.5461 [hep-th].
  
\bibitem{Belopolsky:1996cy}
A. Belopolsky, “De Rham cohomology of the supermanifolds and superstring BRST co- homology,” Phys. Lett. B 403, 47 (1997) [hep-th/9609220]; “New geometrical approach to superstrings,” hep-th/9703183.
 
\bibitem{Belopolsky:1997jz}
A. Belopolsky, “Picture changing operators in supergeometry and superstring theory,” hep-th/9706033.

\bibitem{DHoker:2001kkt}
E. D’Hoker and D. H. Phong, “Two loop superstrings. I. Main formulas,” Phys. Lett. B 529, 241 (2002) [hep-th/0110247]. “II. The Chiral measure on moduli space,” Nucl. Phys. B 636, 3 (2002) [hep-th/0110283]. “III. Slice independence and absence of ambigu- ities,” Nucl. Phys. B 636, 61 (2002) [hep-th/0111016]. “IV: The Cosmological constant and modular forms,” Nucl. Phys. B 639, 129 (2002) [hep-th/0111040]. “V. Gauge slice in- dependence of the N-point function,” Nucl. Phys. B 715, 91 (2005) [hep-th/0501196]. “VI: Non-renormalization theorems and the 4-point function,” Nucl. Phys. B 715, 3 (2005) [hep-th/0501197]. “VII. Cohomology of Chiral Amplitudes,” Nucl. Phys. B 804, 421 (2008) [arXiv:0711.4314 [hep-th]].   

\bibitem{Witten:2012bg}
E. Witten, “Notes On Supermanifolds and Integration,” arXiv:1209.2199 [hep-th]; “Notes On Super Riemann Surfaces And Their Moduli,” arXiv:1209.2459 [hep-th]; “Notes On Holomorphic String And Superstring Theory Measures Of Low Genus,” arXiv:1306.3621 [hep-th].

\bibitem{Donagi:2013spo}
R. Donagi and E. Witten, “Supermoduli Space Is Not Projected,” arXiv:1304.7798 [hep-th]; “Super Atiyah classes and obstructions to splitting of supermoduli space,” arXiv:1404.6257 [hep-th].

\bibitem{Sen:2014pia}
  A.~Sen, ``Off-shell Amplitudes in Superstring Theory", Fortsch.Phys. 63 (2015) 149-188 [arXiv: 1408.0571 [hep-th]].

\bibitem{Friedan:1985ge}
   D. Friedan, E. J. Martinec and S. H. Shenker, “Conformal Invariance, Supersymmetry and String Theory,” Nucl. Phys. B 271, 93 (1986).

\bibitem{Verlinde:1987sd}
     E.~Verlinde and H.~Verlinde, ``Multiloop Calculations in Covariant Superstring Theory",  Phys. Lett. B 192 (1987) 95-102.

\bibitem{Sen:2013oza}
    A. Sen, ``S-duality Improved Superstring Perturbation Theory", JHEP 11 (2013) 029 [arXiv:1304.0458] [INSPIRE].   

\bibitem{Sen:2016gqt}
    A. Sen, ``One loop mass renormalization of unstable particles in superstring theory", JHEP 11 (2016) 050 [arXiv:1607.06500] [INSPIRE].

\bibitem{Verlinde:1986kw}
    E. P. Verlinde and H. L. Verlinde, “Chiral Bosonization, Determinants and the String Partition Function,” Nucl. Phys. B 288, 357 (1987).

\bibitem{Atick:1986rs}
Joseph J. Atick and Ashoke Sen, “Covariant One Loop Fermion Emission Amplitudes in Closed String Theories”, Nucl. Phys. B 10.1016/0550-3213(87)90075-7

\bibitem{Friedan:1985ge}
Daniel Friedan, Emil J. Martinec, Stephen H. Shenker ``Conformal Invariance, Supersymmetry and String Theory", Nucl.Phys.B 271 (1986) 93-165.

\bibitem{Friedan:1986ua}
D. Friedan, S. Shenker ``The Analytic Geometry of Two-Dimensional Conformal Field Theory", Nucl.Phys.B 281 (1987) 509-545; ``The Integrable Analytic Geometry of Quantum String", Phys.Lett.B 175 (1986) 287-296 .

\bibitem{Polchinski:1998rr}
	J.~Polchinski, ``String theory. Vol. 1: An introduction to bosonic string", Cambridge University Press.\\
	J.~Polchinski, ``String theory. Vol. 2: Superstring theory and beyond", Cambridge University Press.
    	      
\bibitem{Sen:SPT21}
A.~Sen, Lectures on ``Superstring Perturbation Theory", LACES, GGI Florence,\\
    	https://www.youtube.com/playlist?list=PLbcx3dKCUhgZv6nW1OiritN2rhfHoVXS4
	
\bibitem{Sen:SPT22}
A.~Sen, Lectures on ``Superstring Perturbation Theory", ICTS Bangalore,\\
	https://www.youtube.com/watch?v=tQguOOdsVHo
    
\bibitem{Igusa1962}
Jun-Ichi Igusa ``On Siegel Modular Forms of Genus Two", American Journal of Mathematics, Vol. 84, No. 1 (Jan., 1962), pp. 175-200.   
\end{thebibliography}
\end{document}